# Trace element geochemistry of CR chondrite metal

Emmanuel Jacquet[1,2], Marine Paulhiac-Pison[1,3], Olivier Alard[4], Anton T. Kearsley[5], Matthieu Gounelle[1,6]

[1]Laboratoire de Minéralogie et Cosmochimie du Muséum, CNRS & Muséum National d'Histoire Naturelle, UMR 7202, 57 rue Cuvier, 75005 Paris, France.

[2]Canadian Institute for Theoretical Astrophysics, University of Toronto, 60 St Georges Street, Toronto, ON, M5S 3H8, Canada.

[3]Ecole Normale Supérieure de Paris, 45 rue d'Ulm 75005 Paris, France.

[4]Géosciences Montpellier, UMR 5243, Université de Montpellier II, Place E. Bataillon, 34095 Montpellier cedex 5, France.

[5]Impacts and Astromaterials Research Centre, Department of Mineralogy, The Natural History Museum, London SW7 5BD, UK.

[6]Institut Universitaire de France, Maison des Universités, 103 boulevard Saint-Michel, 75005 Paris, France.

E-mail: ejacquet@cita.utoronto.ca

## Abstract

We report trace element analyses by laser ablation inductively coupled plasma mass spectrometry (LA-ICP-MS) of metal grains from nine different CR chondrites, distinguishing grains from chondrule interior ("interior grains"), chondrule surficial shells ("margin grains") and the matrix ("isolated grains"). Save for a few anomalous grains, Ni-normalized trace element patterns are similar for all three petrographical settings, with largely unfractioned refractory siderophile elements and depleted volatile Au, Cu, Ag, S. All three types of grains are interpreted to derive from a common precursor approximated by the least melted, fine-grained objects in CR chondrites. This also excludes recondensation of metal vapor as the origin of the bulk of margin grains. The metal precursors presumably formed by incomplete condensation, with evidence for high-temperature isolation of refractory platinum-group-element (PGE)-rich condensates before mixing with lower temperature PGE-depleted condensates. The rounded shape of the Ni-rich, interior grains shows that they were molten and that they equilibrated with silicates upon slow cooling (1-100 K/h), largely by oxidation/evaporation of Fe, hence their high Pd content for example. We propose that Ni-poorer, amoeboid margin grains, often included in the pyroxene-rich periphery common to type I chondrules, result from less intense processing of a rim accreted onto the chondrule subsequent to the melting event recorded by the interior grains. This means either that there were two separate heating events, which formed olivine/interior grains and pyroxene/margin grains, respectively, between which dust was accreted around the chondrule, or there was a single high-temperature event, of which the chondrule margin record a late "quenching phase", in which case dust accreted onto chondrules while they were molten. In the latter case, high dust concentrations in the chondrule-forming region (at least 3 orders of magnitude above minimum mass solar nebula models) are indicated.

## 1. Introduction

Metal has long been recognized an important component of primitive meteorites (Howard 1802; Urey and Craig 1953; Wood 1967; Campbell et al. 2005). Dominated by iron-nickel alloys (chiefly kamacite and taenite), the metal fraction varies between 0 and 70 vol% in chondrites and is one classical petrographic discriminant among different chemical groups (e.g. Brearley and Jones 1998; Scott and Krot 2003; Rochette et al. 2008). Part of this variation is due to oxidation/reduction processes implying exchange of iron between the metal and silicate fraction (Prior 1916; Urey and Craig 1953), but significant variability of the total Fe/Si ratio suggests a further process of metal/silicate fractionation (Urey and Craig 1953; Larimer and Anders 1970).

In thermodynamic calculations, the bulk of iron metal is predicted to condense out of a solar gas at temperatures similar to those of forsteritic olivine (1357 vs. 1354 K at 10 Pa according to Lodders 2003), with higher temperature metallic condensates being enriched in refractory siderophile elements. Indeed, some iron-nickel metal has been occasionally found in amoeboid olivine aggregates (Weisberg et al. 2004) and platinum-group-element-rich nuggets occur in some calcium-aluminum-rich inclusions (Campbell et al. 2005). While this suggests that some chondritic metal may be ultimately the result of nebular condensation, the origin of a significant proportion of it is likely linked to chondrule formation (Campbell et al. 2005), a process as yet involved in obscurity. Indeed, not only do many chondrules contain metal, but the general deficiency of chondrules in siderophile elements relative to CI chondritic abundances (Sears et al. 1996; Hewins et al. 1997; Gordon 2009) suggests that many metal grains now found isolated in the matrix may have originated from chondrules (Uesugi et al. 2008).

Unravelling the origin and history of chondritic metal is thus of undisputable cosmochemical importance, yet a consensus on the subject is not currently at hand. Broadly speaking, possible scenarios regarding metal generation are:

(i) Direct condensation in the protoplanetary disk (e.g. Wood 1963; Campbell and Humayun 2004).
(ii) Recondensation following local evaporation (e.g. Connolly et al. 2001).
(iii) Reduction of iron contained in silicates, be the reducing agent carbon (e.g. Connolly et al. 2001) or nebular gas (e.g. Zanda et al. 1994).
(iv) Desulfurization of sulfides by partial volatilization (e.g. Hewins et al. 1997).

CR chondrites (Weisberg et al. 1993) represent a unique group to make progress on this issue. Metal is abundant (5-8 vol%) and primarily present in type I (low-FeO) chondrules (largely devoid of sulfides) where it forms distinctive outer shells, as though CR chondrules represented an early stage prior to iron loss from chondrules. Indeed some textures seem to be "fossilized" just as metal grains were escaping from chondrules (e.g. Campbell et al. 2005). CR chondrites also largely escaped thermal metamorphism, allowing pre-accretionary features of metal to be preserved, unlike e.g. most ordinary chondrites where metal carries the overprint of the thermal history of the parent body (Afiattalab and Wasson 1980; Kimura et al. 2008), and may comprise the most pristine chondrites known (e.g. Queen Alexandra Range (QUE) 99177 and Meteorite Hill (MET) 00426; see Abreu and Brearley 2010). Pb-Pb ages of CR chondrite chondrules (~4565 Ma) are comparable to other chondrite groups (Amelin et al. 2002; Charles and Davis 2010), as is the Hf-W age of NWA 801 metal (Quitté et al. 2010) although Al-Mg ages seem systematically younger (Kita and Ushikubo 2012). From a chemical standpoint, Co and Ni in CR metal correlate around the solar ratio (Co/Ni = 0.047 (in mass); Lodders 2003), a property only shared by CH and CB chondrites, with which they form the "CR clan" (e.g. Krot et al. 2002), and also by the ungrouped meteorite Acfer 094 (Kimura et al. 2008).

Lee et al. (1992) noted that interior metal grains tended to be richer in Ni than margin grains, which exhibited Ni zoning (inverted U profile). They attributed this to *in situ* reduction on the parent body, where reduction would not have affected interior grains, but the low degree of thermal metamorphism recognized since for these meteorites seem to rule this out (Wasson and Rubin 2010). Weisberg et al. (1993) saw in the Co-Ni correlation an evidence for a condensation scenario, where Ni-richer interior grains represented earlier, higher-temperature condensates than the subsequently accreted margin grains, but Humayun et al. (2002) excluded it e.g. on the basis of Pd/Fe fractionation. Based on Secondary Ion Mass Spectrometer (SIMS) analyses, Connolly et al. (2001) proposed that some margin grains, depleted in refractory siderophile elements, formed by recondensation from surrounding vapor, while others could be interior grains, presumably formed by reduction, in the process of being expelled. Wasson and Rubin (2010) attributed the apparent motion of metal toward the periphery to surface tension effects, similarly to Wood (1963), and hypothesized that margin metal formed a continuous film around the chondrules that beaded upon cooling, while interior grains represented an earlier (unmelted) generation of metal. In their Instrumental Neutron Activation Analyses (INAA), Kong and Palme (1999) found that coarse- and fine-grained metal had similar chemical signatures, suggesting a genetic relationship.

Trace element geochemistry is a potentially powerful tool to shed light on these issues (Connolly et al. 2001; Humayun 2012). The various elements contained in metal present a wide spectrum of volatilities ranging from refractory platinum-group elements (PGE; with the exception of Pd) to moderately volatile Cu, Au, Ag and S, with "main component" siderophile elements (Fe, Co, Ni, Pd) in between (Palme 2008). Likewise, they exhibit a range of geochemical affinities, with highly siderophile elements like PGEs or Ni coexisting with Si, P which become lithophile under oxidizing conditions, and chalcophile Au, Ag, Cu and S. This array of cosmochemical properties could help constrain the origin of metal.

We previously presented trace element analyses using Laser Ablation Inductively Coupled Plasma Mass Spectrometry (LA-ICP-MS) in *silicate* phases in chondrules, including chondrules from the CR chondrites Renazzo and Acfer 187 (Jacquet et al., 2012). In this paper, we present LA-ICP-MS analyses of 66 metal grains in 9 different CR chondrites. With a study of coarse metal grains of Acfer 097 (CR2) by Humayun (2012), as well as further analyses reported in abstract form (Humayun et al. 2002; Humayun et al. 2010; Humayun 2010), this is one of the first applications of this technique to CR chondrite metal. After description of the samples and methods, we will present the results (petrographical and chemical) and discuss the history of metal and the scenario that emerges for CR chondrite chondrules, before concluding.

## 2. Samples and analytical procedures

**Table 1**: Samples used in this study and number of grains analyzed

| Meteorite | Section number | Section type | Origin | Number of metal grains analyzed by LA-ICP-MS | Interior grains | Margin grains | Isolated grains |
|---|---|---|---|---|---|---|---|
| Acfer 187 |  | thick | MNHN | 9 | 6 | 2 | 1 |
| DaG 574 | 2 | thick | MNHN | 2 | 1 | 0 | 1 |
| GRA 06100 | 14 | thick | NASA | 7 | 3 | 2 | 2 |
| GRO 03316 |  | thin | NASA | 6 | 2 | 2 | 2 |
| LAP 02342 | 25 | thin | NASA | 7 | 3 | 3 | 1 |
| LAP 04516 | 8 | thick | NASA | 6 | 2 | 2 | 2 |
| LAP 04592 | 5 | thick | NASA | 6 | 2 | 2 | 2 |
| MET 00426 | 74 | thin | NASA | 9 | 4 | 3 | 2 |
| QUE 99177 | 27 | thin | NASA | 7 | 2 | 2 | 3 |
| Renazzo | 719 | thick | MNHN | 7 | 3 | 2 | 2 |
| Renazzo | NS2 | thick | MNHN | 8 | 2 | 2 | 4 |

MNHN = Museum National d'Histoire Naturelle, NASA = National Aeronautics and Space Administration (Astromaterial Acquisition and Curation Office)

Meteorite names: DaG = Dar al Gani, GRA = Graves Nunatak, GRO = Grosvenor Mountains, LAP = LaPaz Ice Field, MET = Meteorite Hills, QUE = Queen Alexandra Range

Eleven sections of ten different CR meteorites were selected for this study and are listed in Table 1. The sections were examined in optical and scanning electron microscopy (SEM; with a JEOL JSM-840A instrument). Cross-sectional area and perimeters of metal grains were determined using the JMicrovision software (Roduit 2012), allowing a convolution index (CVI) to be calculated (similarly to Zanda et al. 2002) as the ratio between the measured perimeter and that of an equal-area disk. The thus defined CVI is necessarily larger than, or equal to 1 (the result for a circular grain) and is greatest for the most irregular outlines.

Minor and major element concentrations of documented chondrules were obtained with a Cameca SX-100 electron microprobe (EMP) at the Centre de Microanalyse de Paris VI (CAMPARIS), using well-characterized mineral standards. The beam current and accelerating voltage were 10 nA and 15 kV, respectively. The following standards were used: diopside for Si, Ca; $MnTiO_3$ for Mn, Ti; pure iron for Fe; pure cobalt for Co; pure nickel for Ni; apatite for P; pyrite for S; sphalerite for Zn;

$Cr_2O_3$ for Cr. Analyses of olivine, pyroxene and mesostasis co-existing with metal were also carried out (conditions as in Jacquet et al. 2012).

Trace element analyses of selected chondrules were performed by LA-ICP-MS at the University of Montpellier II. The laser ablation system was a GeoLas $Q^+$ platform with an Excimer CompEx 102 laser and was coupled to a ThermoFinnigan Element XR mass spectrometer. The ICP-MS was operated at 1350 W and tuned daily to produce maximum sensitivity for the medium and high masses, while keeping the oxide production rate low ($^{248}ThO/^{232}Th \leq 1\%$). Ablations were performed in pure He-atmosphere ($0.65 \pm 0.05$ l•min$^{-1}$) mixed before entering the torch with a flow of Ar ($\approx 1.00 \pm 0.05$ l•min$^{-1}$). Laser ablation conditions were: fluences ca. 12J/cm² with pulse frequencies of 5 Hz were used and spot sizes of 51 or 77 µm. Each analysis consisted of 3 min of background analyses (laser off) and 1 min of ablation (laser on). Long background times and short analysis times were chosen in order to have a statistically meaningful analysis of a low background noise. Data reduction was carried out using the GLITTER software (Griffin et al. 2008). The external standard PGE-A (Alard et al. 2000), whose long-term reproducibility and applicability to Fe-Ni matrices had been demonstrated by Mullane et al. (2004), was used to calibrate Os, Ir, Ru, Rh, Pt, Pd, Au, Ag, Se, As, Pb, Bi, Te (see also Burton et al. 2012 for Pb ; Lorand and Alard 2010 ; Lorand and Alard 2011 for Te and Bi). Co and Cu were calibrated on the standard steel NIST SRM 1262b. Qualitative (not sufficiently calibrated) data (not reported here) were also collected for S, Cd, Zn, Si, P, W, Re. In all cases, the internal standard was Ni known from EMP analyses. This double standardization allows correction for variations in ablation yield and instrumental drift (Longerich et al. 1996). Minimum detection limit (for background-subtracted signal) is set at 2.3 times the standard deviation of the background (from Poisson statistics), corresponding to a confidence level of 99 % (Griffin et al. 2008).

## *3. Results*

### 3.1 Petrography

In this section we briefly describe the texture and petrographical setting of metal grains in the CR chondrites studied (Fig. 1 and 2). Metal occurs either as isolated grains in the matrix (henceforth "isolated grains") or inside chondrules, either in the interior ("interior grains") or at the surface ("margin grains"). Isolated grains tend to be mostly rounded in shape; some of them could actually be margin grains, with the host chondrule lying outside the plane of the section, as suggested in some cases by some attached silicates. All but one of the chondrules hosting the interior or margin grains analyzed in this study were type I porphyritic chondrules. There is a textural continuum from

*slightly melted*, fine-grained chondrules with convoluted outlines to *highly melted*, round chondrules, with *medium melted* chondrules being intermediate (see Fig. 2a-c; Zanda et al. 1993, 2002): in the slightly melted objects (which we also call "chondrules" following Zanda et al. (2002) and Hutchison (2004)), metal is fine-grained (of the order 10 µm in size, too small to be analyzed by LA-ICP-MS) and peppers its host more or less uniformly. With increasing melting degree, metal grains tend to coalesce, as evidenced by the amoeboid shape of the resulting larger grains (of order 100 µm in size), and tend to concentrate around the margin. In the highly melted chondrules, interior grains are rounded, although margin grains, which never form a complete, continuous shell around the chondrules, are mostly amoeboid in shape. Pyroxene (enstatite, with augite overgrowths) in most type I chondrules is concentrated near the border, while olivine phenocrysts, sometimes meeting in triple junctions (although full-fledged granoblastic olivine aggregates like those described by Libourel and Krot (2007) are very rare), dominate the interior, and margin metal grains are usually included in the pyroxene-rich layer (see X-ray map in Fig. 1). Type I chondrules also typically have an outermost igneous rim similar to the the least-melted chondrules (e.g. Fig. 2d). Metal in the single type II (high-FeO) chondrule studied (in LAP 02342) is associated with troilite (see Fig. 2j). Sulfides are rare in type I chondrules (although they are more abundant in the matrix) and mostly occur as thin partial rinds around chondrules or metal grains (Fig. 2f).

Metal grains appear generally homogeneous, although our BSE observation may easily miss their frequent polycrystallinity noted by Wood (1967). Several interior grains, most prominently in GRO 03116, exhibit Ni-rich (15-30 wt% Ni) worm-shaped exsolutions (of the order of 10 µm in size) apparent in low-brightness, high-contrast back-scattered electrons (BSE) images (see Fig. 2h-i), a texture akin to the "cellular plessite" of Buchwald (1975) or the "type III plessite" of Massalski et al. (1966) which may reflect the polycrystallinity of the grains. All grains investigated in GRA 06100, regardless of petrographical setting, show a coarser intergrowth, with Ni-rich areas being more isolated and more blocky in shape (see Fig. 2j). Such textures are investigated in more detail by Briani et al. (submitted; 2010; see also Abreu et al. (2012)). No plessitic texture was encountered in the observed grains in Acfer 187, LAP 04516, MET 00426 and QUE 99177. Micron-size chromite and larger rounded silicate (pyroxene and/or silica) inclusions sometime occur in metal (see also Zanda et al. 1994). One metal/sulfide association in a type II chondrule, similar to those described by Schrader et al. (2010) was also analysed (Fig. 2l).

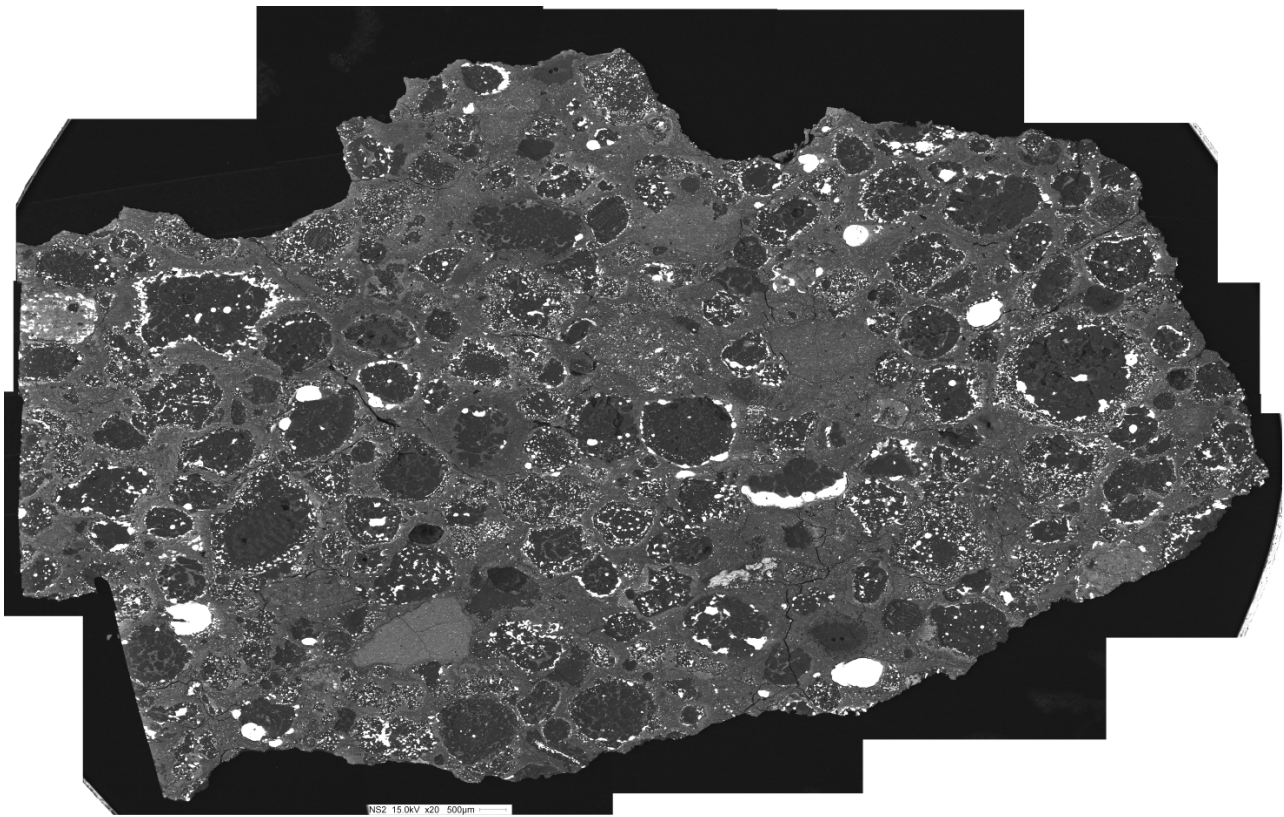
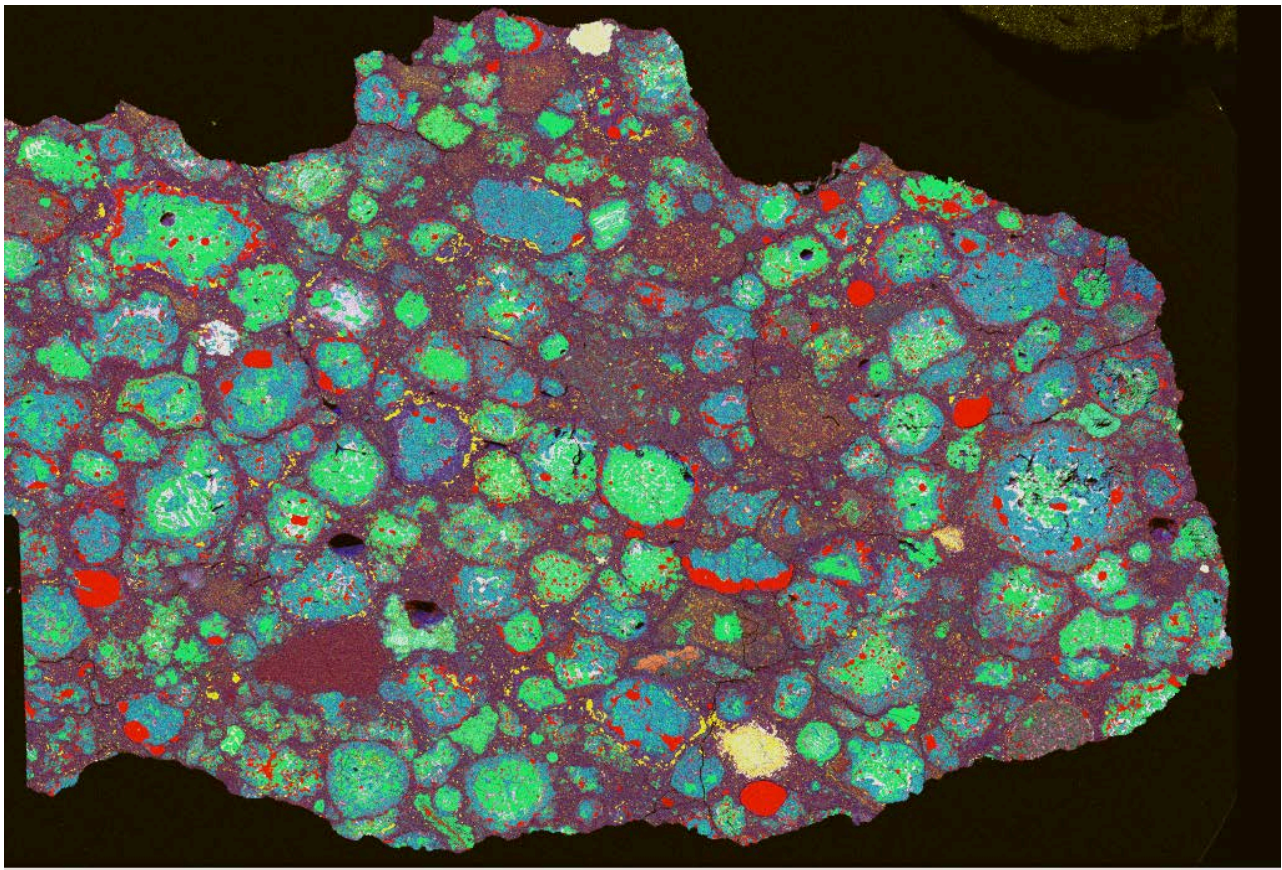

**Figure 1**: Maps of section NS2 of Renazzo. *(Top)* Back-scattered electron (BSE) map. *(Bottom)* X-ray map, with Al in white, Mg in green, Si in blue, Ca in yellow and Fe in red. In this color coding, forsteritic olivine appears green, enstatitic pyroxene is blue and Fe-Ni metal is red.

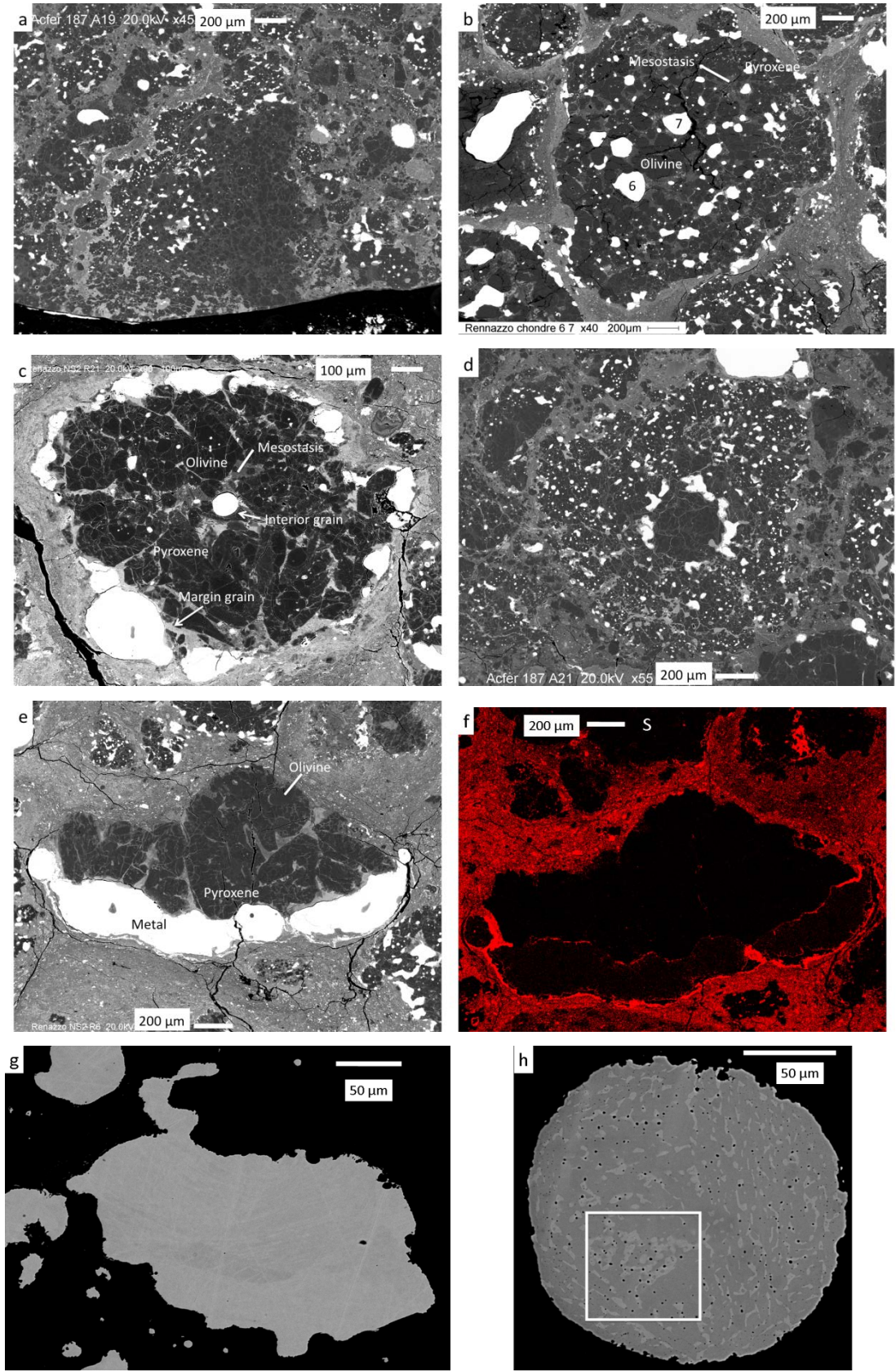

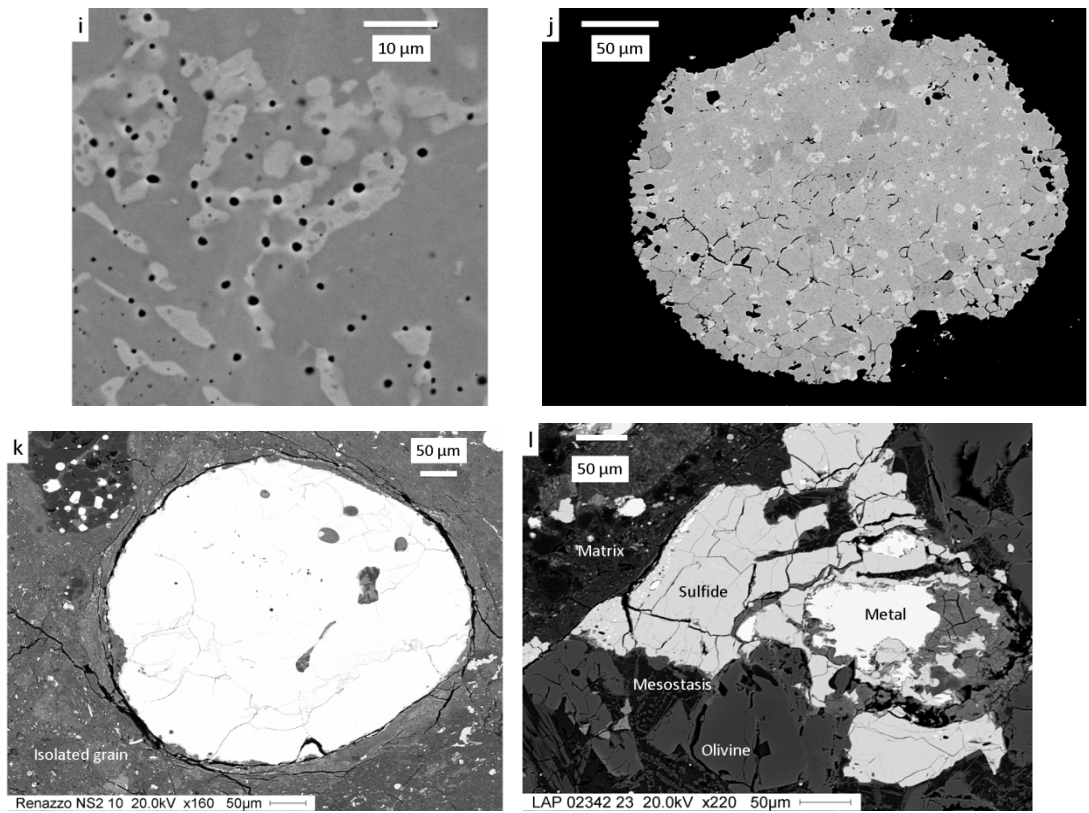

**Figure 2**: Images of metal grains in CR chondrites (BSE images unless otherwise noted). (a) Slightly melted chondrule A19 in Acfer 187, with irregular morphology and fine grain size (<50 µm). Metal is concentrated on the left-hand-side and coarsens in the interior. (b) Medium melted chondrule in Renazzo. The largest metal grains (numbered 6 and 7) are about 200 µm in diameter (the size of the coarsest olivine phenocrysts) and of roughly equant shape but smaller grains are amoeboid. Metal is roughly uniformly distributed throughout the chondrule. Enstatite occurs near the margin. (c) Highly melted chondrule R21 in Renazzo, of ellipsoidal shape. Compared to the previous objects, the chondrule interior is depleted in metal, with a prominent 100 µm diameter round metal grain near the center. Coarse (100-400 µm) amoeboid metal grains rim the chondrule. Pyroxene invades the olivine-dominated chondrule interior, with a mesostasis patch around the center metal grain. No igneous rim is visible around the chondrule but hydrous alteration products are evident outside the metal rim. (d) The chondrule A21 in Acfer 187 contains a 400 µm diameter core with a metallic shell similar to common highly melted chondrules such as (c) and is surrounded by an irregular 200-600 µm thick igneous rim. Large (>200 µm) elongate enstatite crystals poikilitically enclose olivine grains throughout the chondrule. Gray areas, in veins or around metal grains, are terrestrial alteration products. (e) Porphyritic pyroxene chondrule fragment R6 in Renazzo. Metal rim the unbroken surface almost continuously (3 distinct (in 2D) metal grains over 2 mm, with thickness of 200-300 µm). (f) X-ray map of the same for sulfur. S occurs mainly in the matrix, but forms a rind (<20 µm thick) around the metal grains which extends inside the chondrule. (g) Margin grain #19 in MET 00426 viewed in high-contrast, low-luminosity to show its featureless interior. (h) Interior grain #1 in GRO 03116 exhibiting a plessitic texture. The outlined area is enlarged in (i). (i) Zoom in grain #1 in GRO 03116. Brighter areas are Ni-rich (15-30 wt%), the gray host has 7-9 wt% Ni. Black micron-sized inclusions are mainly chromite with accessory phosphate. (j) Interior grain #3 in GRA 06100, exhibiting a relative coarse intergrowth between Ni-poor and Ni-rich metal. (k) Rounded, isolated grain #10 in Renazzo. Some silicate inclusions occur (with pyroxene in light gray and silica in dark gray). (l) Metal grain #23, with corroded morphology, included within troilite in a type II chondrule in LAP 02342. Some rust due to weathering appears on the right.

## 3.2 Mineral chemistry

**Table 2**: Averaged composition for the three petrographical settings

|    | Interior grains | | Margin grains | | Isolated grains | |
|----|------|------|------|------|------|------|
|    | Mean | 1 σ  | Mean | 1 σ  | Mean | 1 σ  |
| P  | 1738 | 161  | 1615 | 157  | 1566 | 194  |
| Fe | 927779 | 4839 | 944882 | 3006 | 944622 | 1410 |
| Co | 3095 | 197  | 2531 | 164  | 2621 | 108  |
| Ni | 69236 | 4369 | 53591 | 2754 | 54770 | 1162 |
| Cu | 49.2 | 3.7  | 69.6 | 11.3 | 35.8 | 11.8 |
| As | 2.64 | 0.30 | 3.37 | 0.43 | 2.68 | 0.40 |
| Os | 3.65 | 0.78 | 2.74 | 0.57 | 2.73 | 0.23 |
| Ir | 3.01 | 0.57 | 2.39 | 0.47 | 2.42 | 0.18 |
| Ru | 3.79 | 0.55 | 3.09 | 0.49 | 2.73 | 0.18 |
| Rh | 0.70 | 0.10 | 0.63 | 0.13 | 0.56 | 0.04 |
| Pt | 4.91 | 0.74 | 4.04 | 0.73 | 3.51 | 0.23 |
| Pd | 3.43 | 0.23 | 2.56 | 0.18 | 2.57 | 0.11 |
| Au | 0.28 | 0.15 | 0.23 | 0.03 | 0.19 | 0.02 |
| Ag | 0.14 | 0.01 | 0.25 | 0.07 | 0.27 | 0.01 |

All values reported are in ppm. Values are from LA-ICP-MS analyses except Fe and Ni (the internal standard). One sigma is the error on the mean (that is, the standard deviation divided by the square root of the number of data).

Chemical data on the analyzed grains are reported in the Electronic Annex and averages over each petrographical setting are shown in Table 2. There is little apparent difference between the studied meteorites—although the limited number of analyses (6-15) per meteorite invites caution—, though EMP analyses of GRA 06100 metal show low Cr contents (< 0.2 wt%) and P is below detection (<0.03 wt%), compared to 0.1-0.3 wt% P and 0.1-0.6 wt% Cr in other CR chondrites. We shall thus discuss the data from the nine CR chondrites studied collectively and simply distinguish the metal grains according to petrographical setting (*viz.* interior, margin or isolated grain). A plot of Co vs Ni is presented in Fig. 3a: we find again the positive correlation around the solar ratio commonly deemed characteristic of CR chondrites (Weisberg et al. 1993; Krot et al. 2002), with interior grains being richer in Ni (4.3-13.6 wt% Ni) than margin and isolated grains (3.3-6.3 wt%; excepting margin grain #19 of MET 00426, with 10.2 wt% Ni). For individual chondrules, the only exceptions to that rule (by <0.5 wt% Ni), at face value, are two medium melted chondrules (one in LAP 02342, with grains #3-6; one in MET 00426, grains #15-18) where "interior" and "margin" grains are not clearly distinguished texturally and one highly melted, rimless chondrule in Renazzo (where the margin grain #4 is the only coarse (~300 µm) margin grain visible in the section for this chondrule and morphologically so similar to interior grain #3 that it may actually be viewed as a migrated interior grain). The Co vs Ni correlation has however significant scatter (as is apparent in similar plots by Connolly et al. 2001; Wasson and Rubin 2010; Krot et al. 2002) and shows a distinct tendency for Ni-rich grains to have a subsolar Co/Ni ratio, as previously noted by Wasson and Rubin (2010). It is worth noting that, based on Figure 5 of Krot et al. (2002), Co-Ni correlation in metal is weaker in CR chondrites than in CH and CB chondrites, possibly suggestive of different geneses. Fig. 3b displays a plot of Ni as a function of the convolution index (CVI) defined in the "Samples and analytical procedures" section 2. It is remarkable that the most Ni-rich grains tend to have rounded shapes: With the exception of one coarse margin grain (#19) in MET 00426 and the single type II chondrule metal grain analyzed (#23 in LAP 02342), all the grains containing more than 8 wt% Ni have a CVI < 1.2. For CVI > 1.5, the Ni contents tend to cluster around the 5.7 wt% expected from a solar metallic condensate (Kelly and Larimer 1977) and the Co/Ni ratio deviates least from the solar value. In a plot of Ni vs grain size (Fig. 3c), these interior Ni-rich grains are found among the *smaller* grains, but margin and isolated grains show a shallow *positive* correlation of Ni with size. No correlation between Ni content of metal and (oxidized) Fe content of olivine is apparent in Fig. 3d, consistent with the analyses of Wasson and Rubin (2010), but there is a tendency for Fe content of olivine to be lower for the chondrules containing the more rounded metal grains (Fig. 3e; disregarding grains with CVI > 1.4), consistent with Zanda et al. (2002) although it must be cautioned that Zanda et al.'s convolution indices were calculated for the whole chondrule shapes rather than metal grains' shapes.

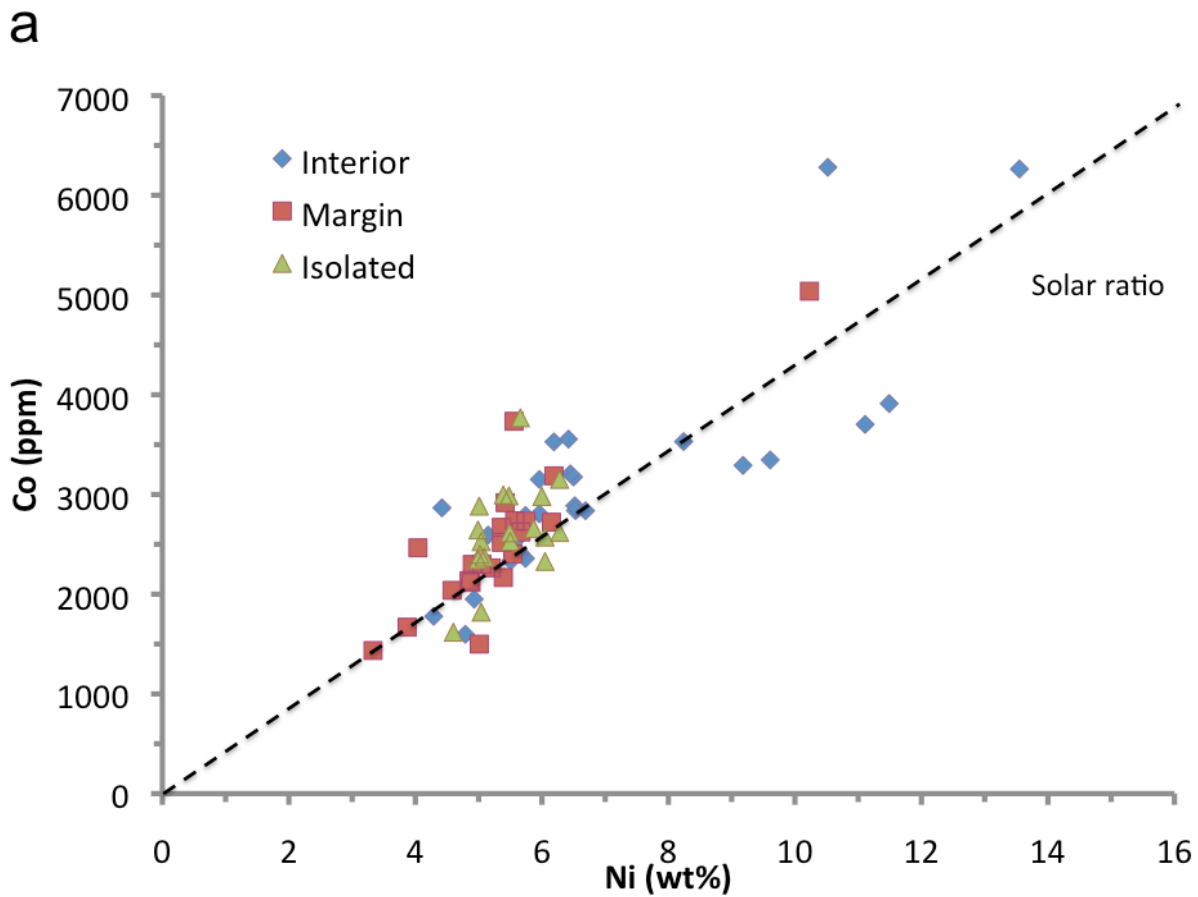

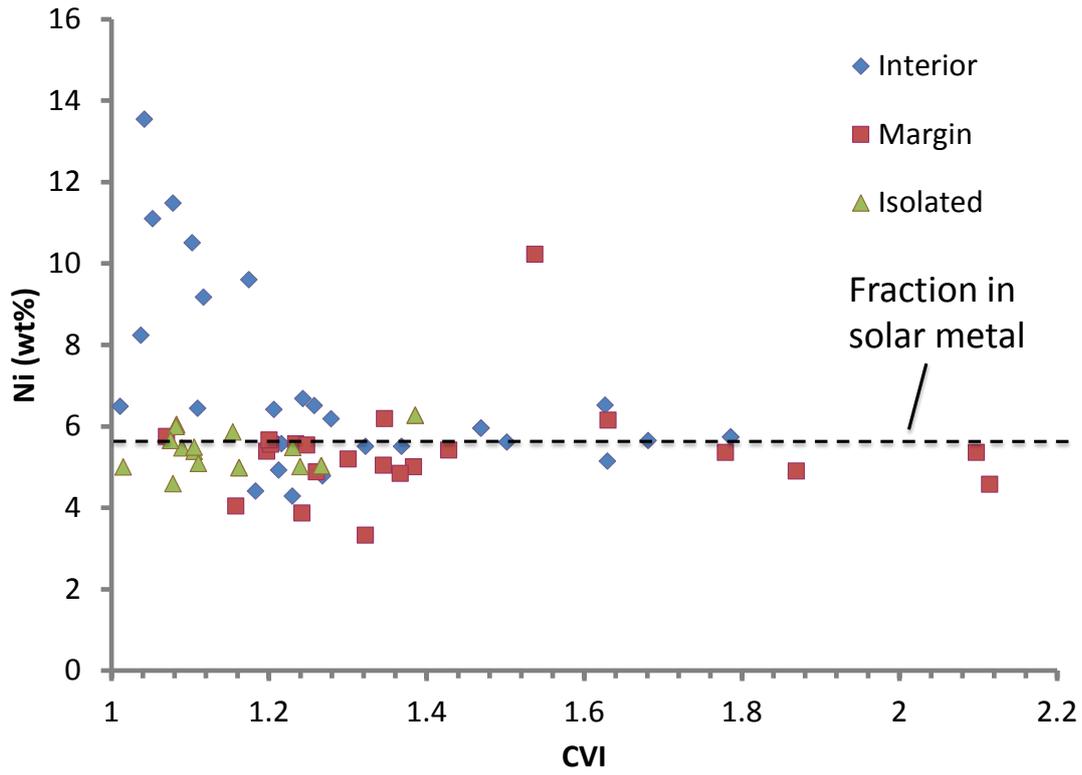

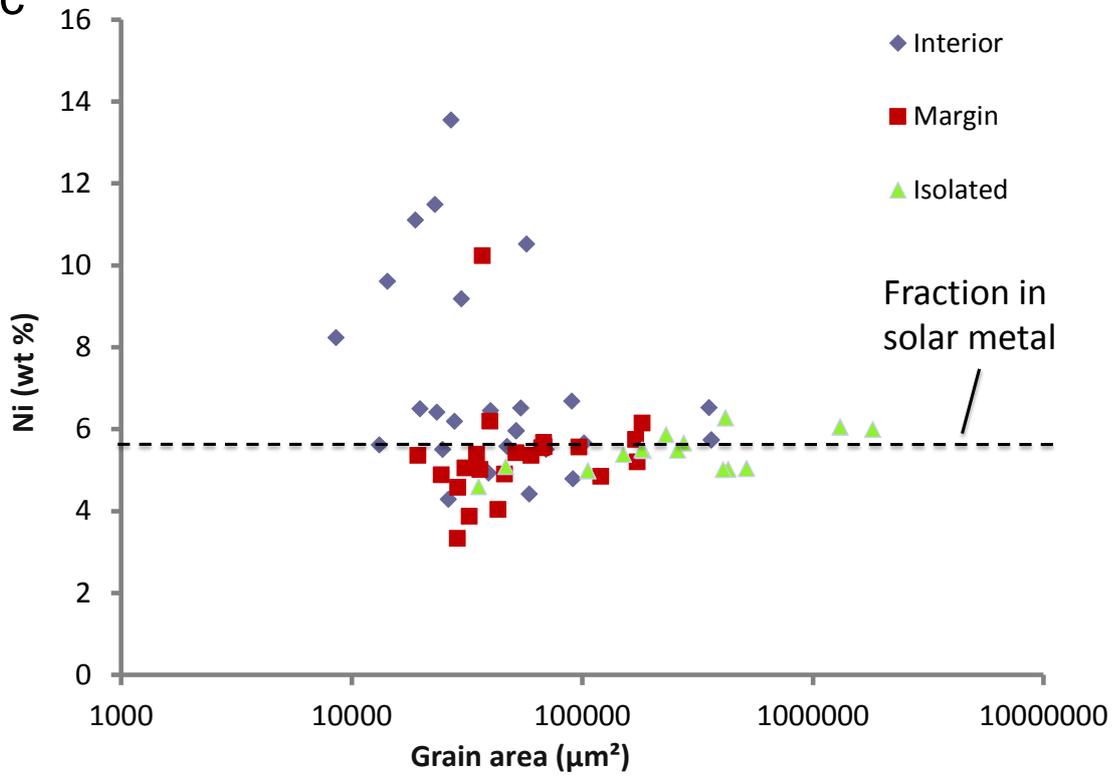

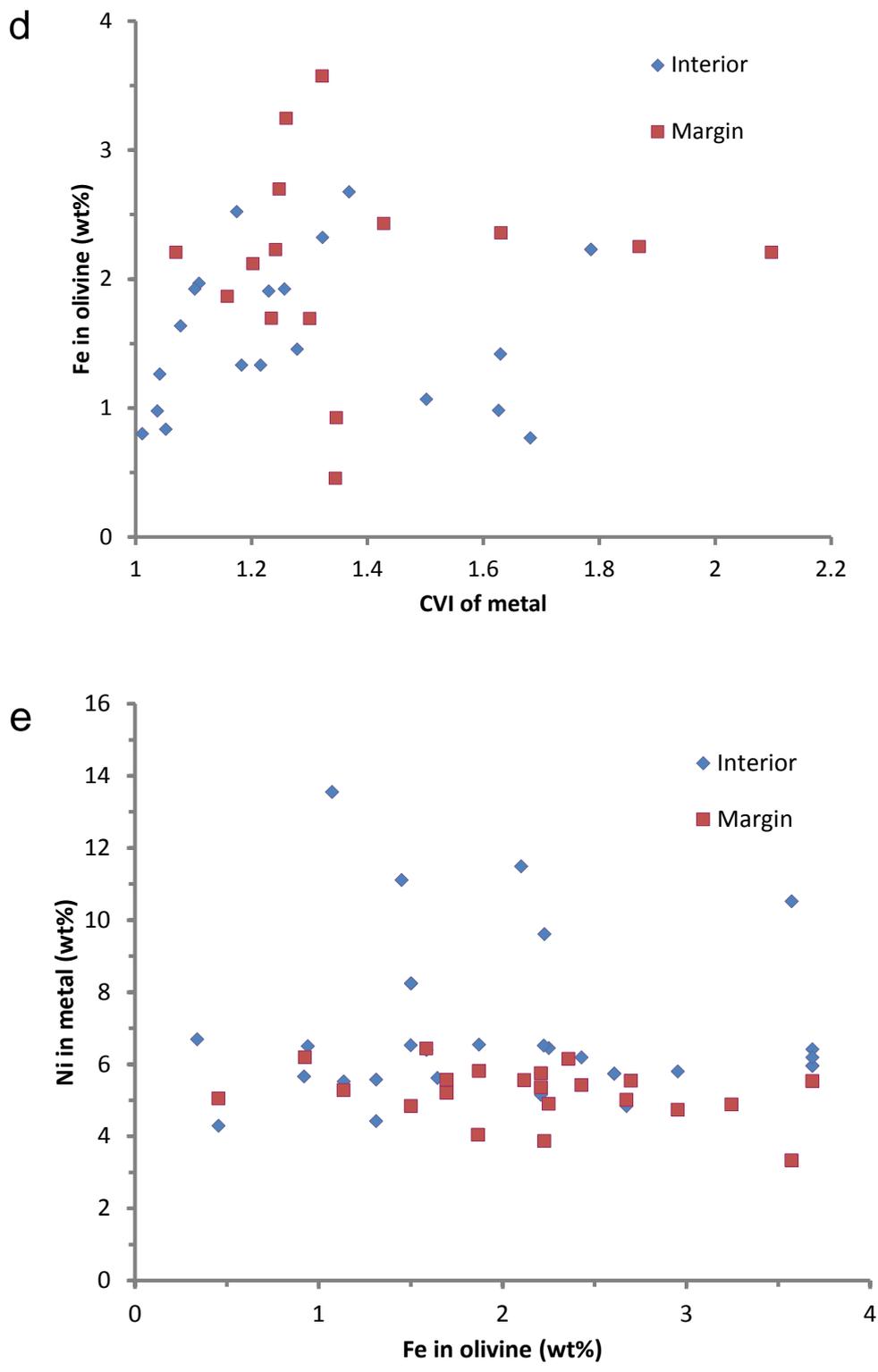

**Figure 3:** *(a)* Plot of Co vs Ni (Co from LA-ICP-MS analyses, Ni from EMP analyses) in metal grains color-coded according to textural setting. The dashed line indicates the solar ratio. *(b)* Plot of Ni as a function of convolution index (CVI, i.e. the ratio between the perimeter and that of the equal-area circle) of the metal grains. The dashed line indicates the 5.7 wt% content expected from a solar metal condensate (Kelly and Larimer 1977). *(c)* Plot of Ni as a function of grain cross-sectional area. *(d)* CVI of interior metal vs. Fe content of associated olivine. *(e)* Plot of Ni in metal and Fe content in chondrule olivine for interior and margin grains.

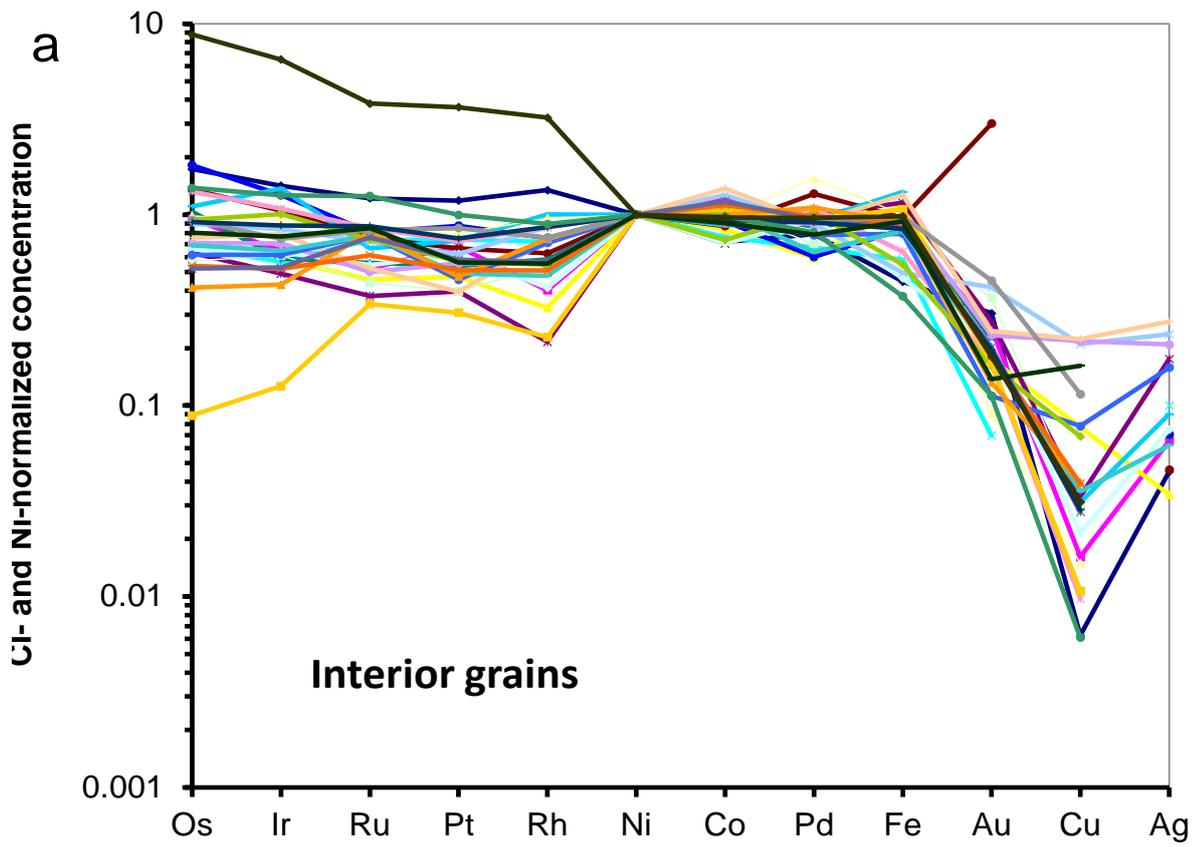
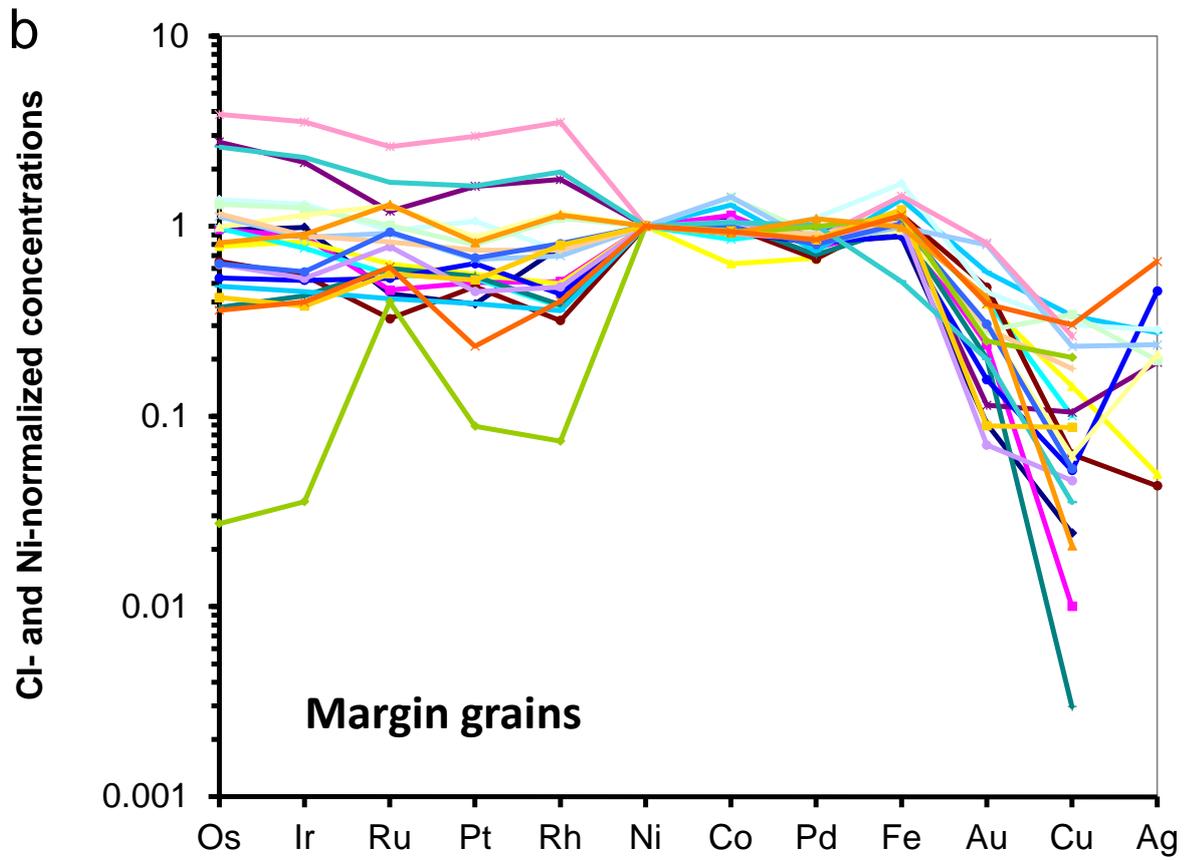

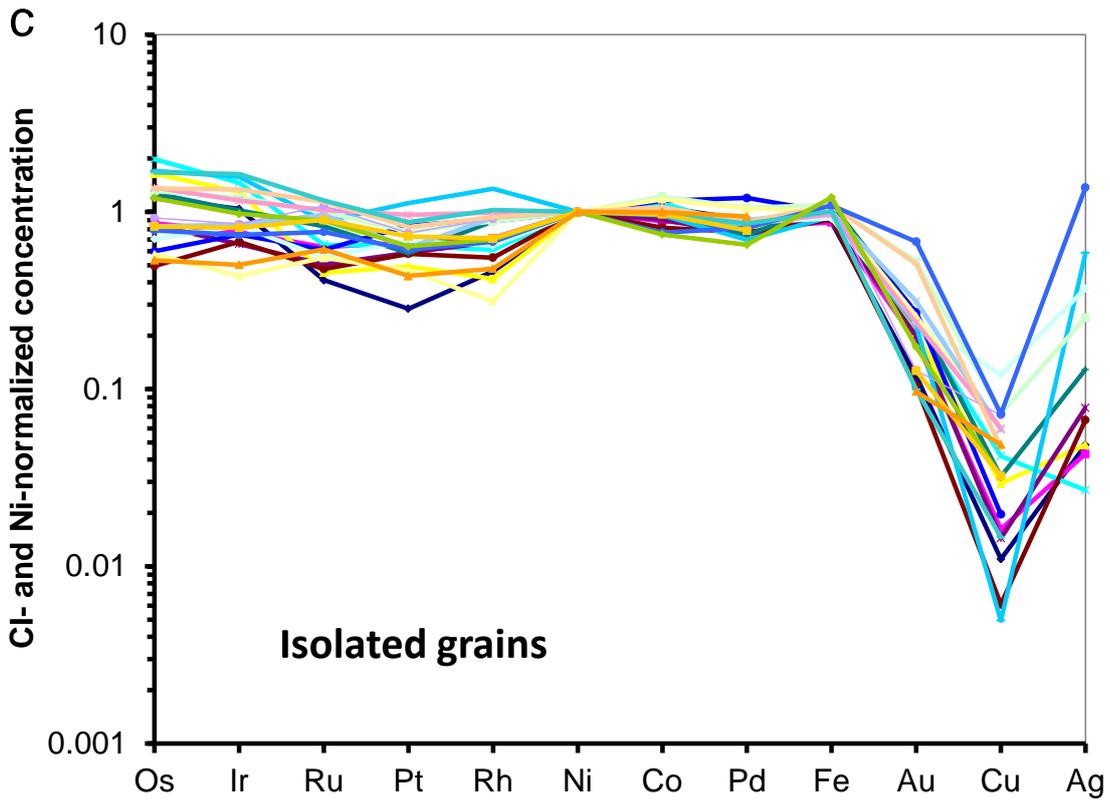

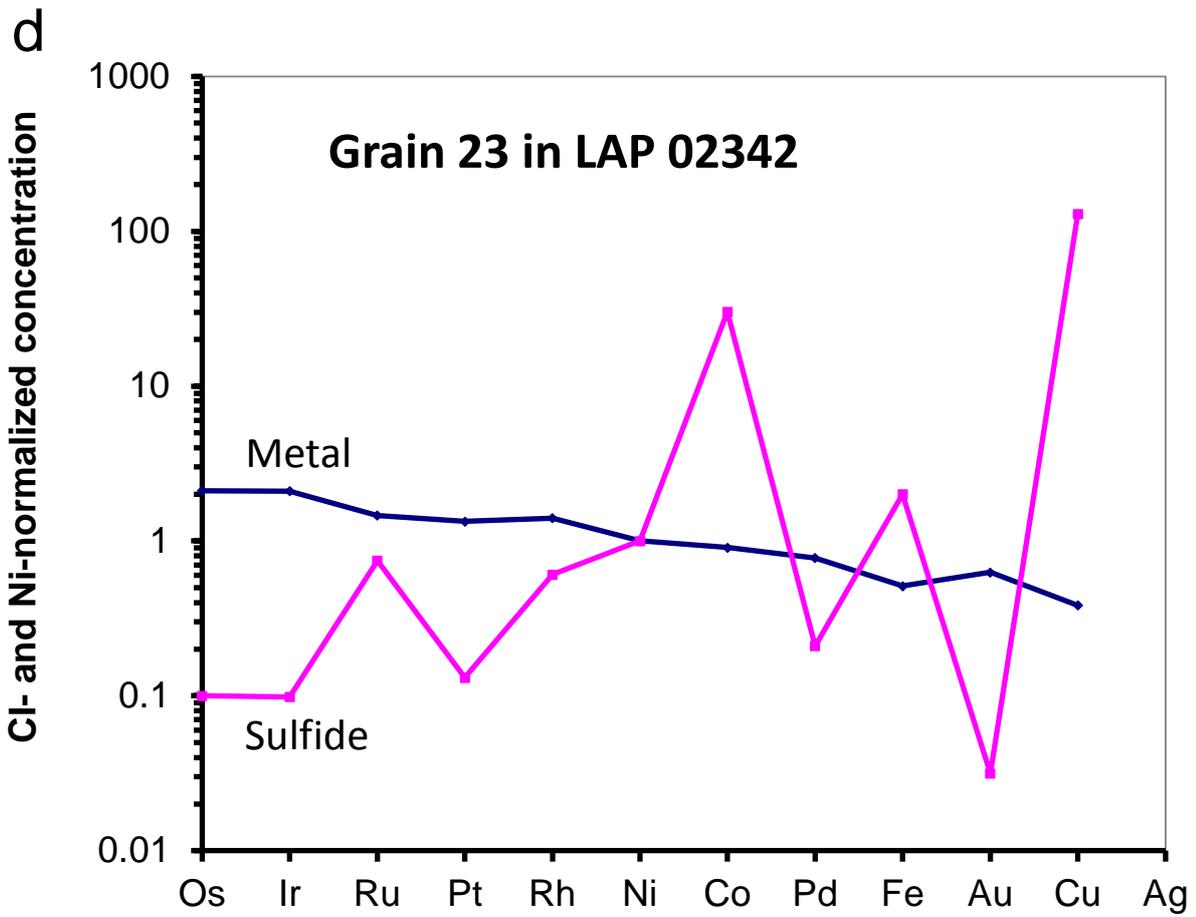

**Figure 4:** Ni- and CI-normalized siderophile element pattern of metal grains, with elements arranged according to decreasing 50 % condensation temperature (Campbell et al. 2003). (a) Interior grains (b) Margin grains (c) Isolated grains (d) Metal (blue)/sulphide (pink) association in the type II chondrule analyzed in LAP 02342. Each color represents one metal grain.

Data for all siderophile elements are presented in CI- and Ni-normalized spider-diagrams in Fig. 4 and averages for each petrographical setting are shown in Fig. 5. All three settings show very similar patterns, consistent with the chemical similarity between fine- and coarse-grained metal observed by Kong and Palme (1999). Generally speaking, refractory elements are unfractionated relative to Ni and agree with literature data (Kong and Palme 1999; Kong et al. 1999; Connolly et al. 2001; Humayun et al. 2002; Humayun et al. 2010; Humayun 2010). We do not see the tendency for margin grains to be depleted in refractory PGEs relative to Ni compared to interior grains reported by Connolly et al. (2001) in their SIMS analyses, which may be due to limited sampling, as this trend is not apparent in the LA-ICP-MS data by Humayun et al. (2010) and Humayun (2010) either. Significant fractionations (enrichments/depletions) of refractory elements are occasionally encountered: if we arbitrarily consider as "anomalous" a grain whose Ni-normalized concentration of any of the refractory element differs by a factor of more than 3 from the average composition, we find that 5 out of 28 interior grains (all from medium melted chondrules), 3 margin grains out of 22 and 1 isolated grain out of 19 are anomalous in this sense, hence a proportion of about 15 %, with isolated grains appearing least variable overall (Fig. 5). Main component siderophile elements (Ni, Co, Pd, Fe) show more limited variation than their refractory counterparts. Moderately volatile elements Au, Cu and Ag are depleted by typically one order of magnitude, with Ag being less depleted than Cu ($(Ag/Ni)_N$ = 0.03-1.4 vs. $(Cu/Ni)_N$ = 0.003-0.3), with margin grains tending to be more enriched. However, this latter trend is not very pronounced (especially in terms of absolute concentrations instead of Ni-normalized abundances), and even the margin grains show lower volatile elements concentration lower than the bulk metal analyses of Kong et al. (1999) (Fig. 5). This may be explained by the concentration of these elements primarily near the rims of these grains (Humayun 2012) whereas our LA-ICP-MS spots mainly targeted the cores of the grains.

The single type II chondrule metal grain analyzed shows a shallower, albeit smooth volatile element depletion trend, with $(Cu/Ni)_N$ = 0.4, while the associated sulphide displays a more saw-tooth-shaped pattern (see Fig. 4d).

Concentrations for Pb, Bi, Te and Se are generally below detection (typically <0.1, 0.01, 1 and 30 ppm, respectively).

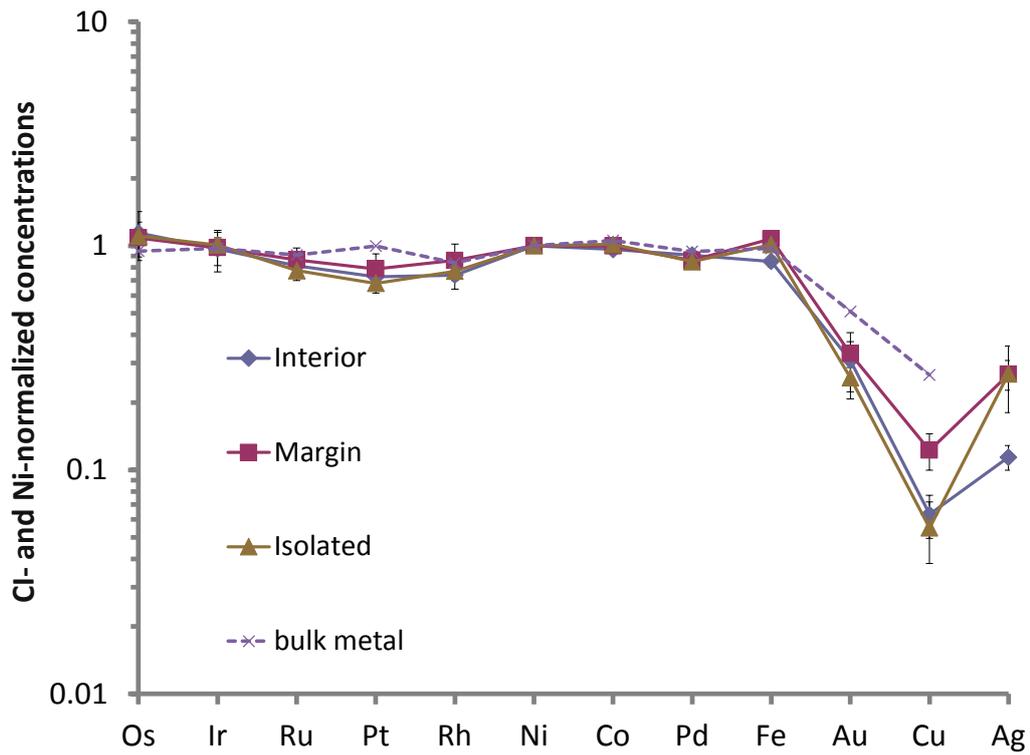

**Figure 5:** Ni- and CI-normalized siderophile element pattern, with elements arranged according to decreasing 50 % condensation temperature (Campbell et al. 2003), for average composition of interior, margin and isolated grains. Bulk metal composition is taken from Kong et al. (1999). Error bars are one standard error of the mean.

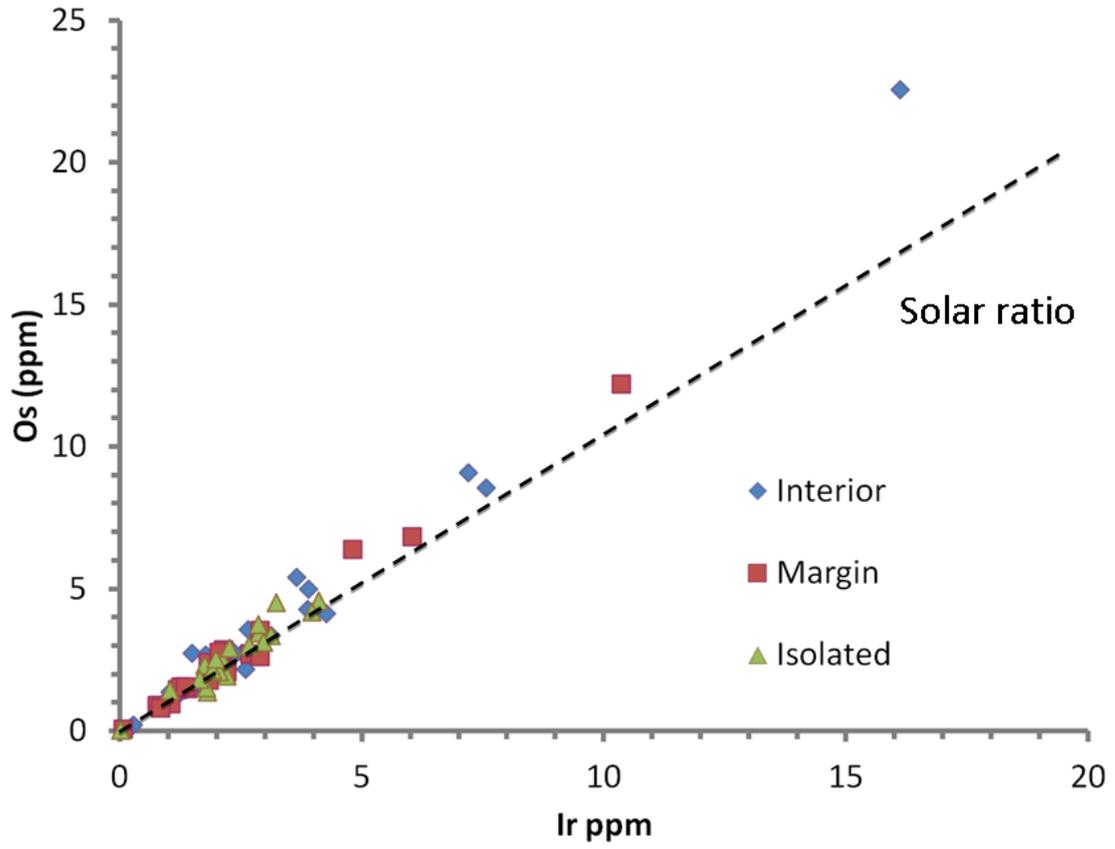

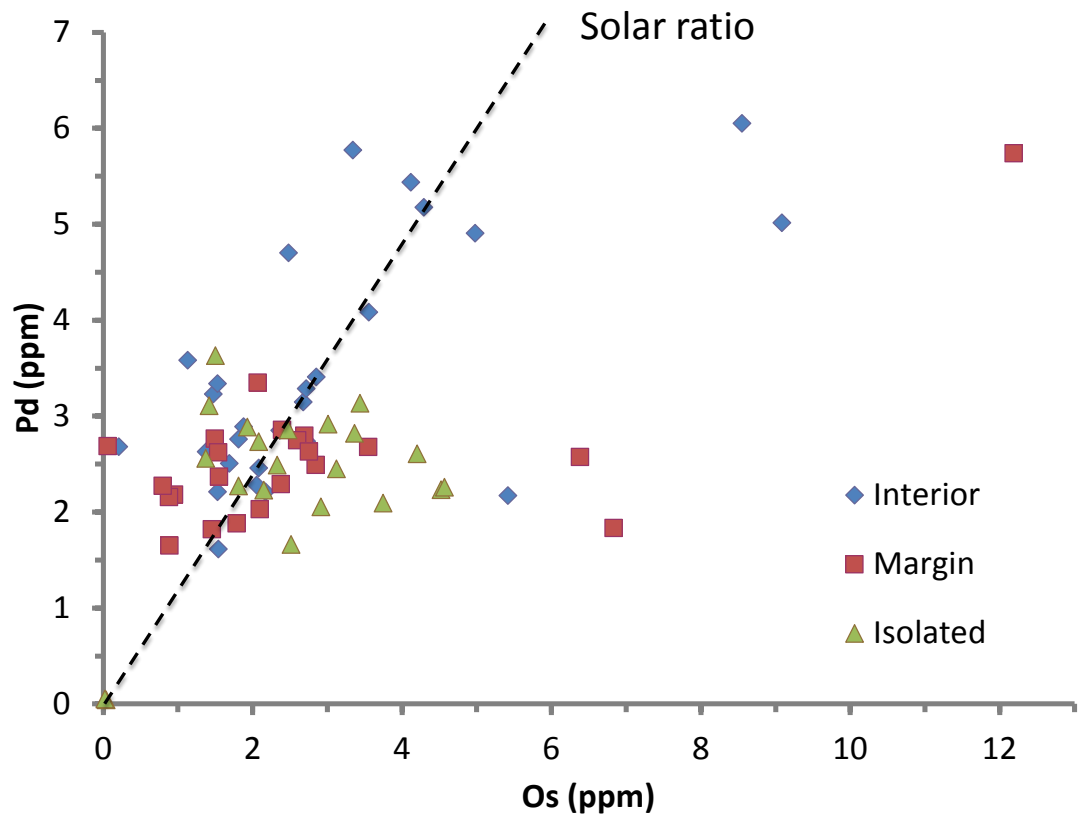

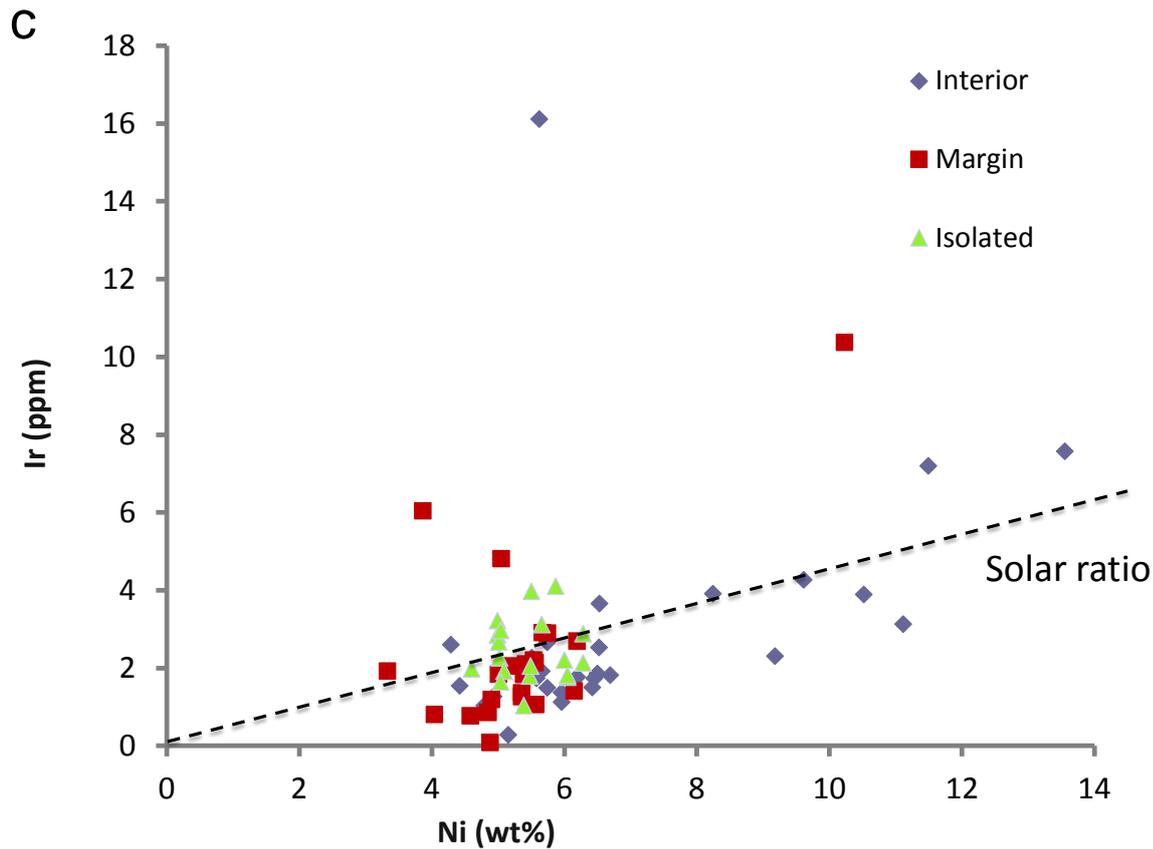

**Figure 6:** *(a)* Plot of Os vs Ir in metal grains in CR chondrites. A strong positive correlation consistent with the solar ratio (dashed line) is apparent. *(b)* Plot of Pd vs. Os. The two elements co-vary (at least for interior grains) but with considerable scatter. *(c)* Plot of Ir vs. Ni in metal grains in CR chondrites. Likewise, there is covariation mostly for interior grains but significant scatter.

## *4. Discussion*

### 4.1 Metal texture: parent body versus primary processes

Before discussing implications pertaining to metal and chondrule formation, an evaluation of parent body effects is in order. Although estimated shock stages of CR chondrites are low (S1 or S2; e.g. Brearley and Jones 1998; Bischoff et al. 1993), thermal metamorphism or aqueous alteration could have left an imprint on CR metal.

We recall from the "Petrography" subsection 3.1 that most metal grains are featureless, but a few interior grains exhibit exsolutions (in particular in GRO 03116), and all grains (irrespective of petrographical setting) of GRA 06100 show an intergrowth of Ni-rich and Ni-poor metal. Kimura et al. (2008) reported that all grains of Acfer 094 and most grains in Semarkona (LL3.0) were featureless, a feature they ascribed to a martensitic transformation during rapid cooling following chondrule formation, whereas higher subtype ordinary chondrites and CO exhibited exsolutions, which they interpreted as resulting from parent body metamorphism. Building on that work, Briani et al. (2010; submitted) concluded that most CR chondrites largely escaped metamorphism (which would have induced some exsolution within the metal grains), but GRO 03116 and GRA 06100 were likely shocked as also suggested by the low hydration state of their matrix and the Raman properties of their organic matter (see also Abreu (2011); Abreu et al. (2012)). The low abundance of P and Cr of GRA 06100 relative to other CR chondrites could be a metamorphic effect as well (Kimura et al. 2008). At the other extreme, it is perhaps no coincidence that no plessitic texture was observed in grains from QUE 99177 and MET 00426, which Abreu and Brearley (2010) ranked at subtype 3.00. The preservation of the Ni-Co correlation, otherwise seen only in the type 3.00 ungrouped Acfer 094 meteorite (Kimura et al. 2008) and, to some extent, in unheated or weakly heated CM chondrites (Kimura et al. 2011), would be consistent with such a pristinity, as upon equilibration, Co partitions preferentially into Ni-poor kamacite.

Thus, most metal grains in CR chondrites would provide essentially pristine records of chondrule formation. Interestingly, the martensitic structure of Ni-rich (>7 wt% Ni) metal yields a lower limit on the cooling rate experienced by CR chondrite chondrules: according to Reisener et al. (2000) and Reisener and Goldstein (2003), martensitic transformation of P-bearing (>0.1 wt%) taenite requires cooling rates >> 0.2 K/h (around 900 K). This is compatible with the 0.5-50 K/h range (at $1473 \pm 100$ K) estimated by Humayun (2012) from the diffusion of Cu and Ga in coarse metal grains in Acfer 097 (CR2), as well as the upper limit of order 10 K/h (during olivine crystallization) inferred from our trace element studies on olivine in CR chondrites (Jacquet et al., 2012).

The origin of occasional plessitic textures in CR chondrites other than GRA 06100 and GRO 03116 would warrant further study. If such textures are to be interpreted as a result of incipient metamorphism (to a lower degree than GRA 06100 or GRO 03116) as suggested by Lee et al. (1992) for Renazzo, the fact that they occur in some grains and not others would indeed require explanation. A possibility could be that inasmuch as those grains are often interior grains, their higher Ni content could allow them to enter the $\alpha + \gamma$ field of the Fe-Ni phase diagram more frequently. However, there appears to be little regularity in this respect in the sense that there are featureless metal grains more nickeliferous than plessitic-textured ones. We cannot thus exclude that some plessitic textures were actually inherited from the chondrule formation process itself, with the metal grains in question lying at the lower end of the array of cooling rates experienced by CR chondrite metal (around 900 K). A dedicated metallographic study of metal grains in CR chondrites would certainly be desirable to settle this issue. Whatever that may be, it is clear that most metal grains in this meteorite group were not significantly modified by parent body processes and thus can be used as indicators of the chondrule-forming process.

## 4.2 Vapor fractionation and chondrule precursors

The systematic depletion in volatile siderophile elements of CR chondrite metal (Fig. 4, 5) hints at vapor fractionation processes. We wish to investigate them here in more detail, and in particular find out whether this depletion arose during chondrule formation, or was inherited from the precursors.

The similarity of the Ni-normalized patterns between interior, margin and isolated grains appears inconsistent with a scenario where margin metal formed by recondensation of evaporated metal during chondrule formation, as proposed by Connolly et al. (2001) for one would then expect margin grains to be depleted in refractory siderophile elements relative to interior grains. In particular, Fig. 6 shows that, relative to margin grains, interior grains are enriched (in absolute concentrations) in Os, Ir, both refractory PGEs, but also in Pd, a moderately volatile one, contrary to what would be expected had vapor fractionation been the sole relevant process (see also Humayun et al. 2002). Nonetheless, enrichment in volatile elements of margin grains relative to interior grains would be consistent with *some* recondensation having affected the trace element budget of margin grains, yielding the outside-in diffusion profiles of Cu and Ga studied by Humayun (2012). (It would be difficult to alternatively envision a partial loss of siderophile volatile element from the interior grains *in situ* since the relatively reducing conditions would have prevented them from leaving the grains and entering the silicate phase in the first place (Campbell et al. 2005)). A petrographical manifestation of this could be the sulfide rinds sometimes encountered around metal grains (e.g. Fig. 2e-f), bearing in mind that sulfur originally in the ambient gas may have been first dissolved in chondrule melt (Marrocchi and Libourel 2012; see also Lauretta et al. 1996). We concur with the physical argument of Wasson and Rubin (2010) that inasmuch as the surface area (inversely proportional to size for a given mass fraction) offered by fine grains for recondensation would dominate that of chondrules, recondensation onto the latter would not have yielded their thick metal mantles, although it might account for the shallow trend of decreasing Ni content with decreasing size (and thus increasing surface/volume ratio) exhibited by margin and isolated metal grains (Fig. 3c).

We do not favor either a derivation of interior or margin grains from volatilization of sulfur from preexisting sulfides (e.g. Campbell et al. 2005), based on the Ni content of the metal (>3 wt% versus e.g. <0.3 wt% in metal from the Camel Donga eucrite inferred to have formed in this manner by Palme 1988), the largely unfractionated nature of refractory siderophile elements compared to the pattern in sulfide associated with metal grain #23 in LAP 02342 (Fig. 4d), and the similar and low (<< 0.1 wt%) S contents of interior and margin grains.

The similarity of volatile element depletion patterns exhibited by the grains regardless of petrographical setting suggests that they were essentially inherited from the chondrule precursors.

The paucity of granoblastic olivine aggregates (GOA; Libourel and Krot 2007) in CR chondrites (Jacquet et al., 2012), suggests that CR chondrite chondrule precursors were not, in general, GOAs, and were likely relatively fine-grained, and the triple junctions that occasionally occur in chondrules were acquired during prolonged heating (Wasson and Rubin 2010; Whattam et al. 2008). Following Zanda et al. (1993, 2002), we suggest that these precursors are best approximated by the slightly melted objects mentioned in the "Petrography" subsection 3.1 (see Fig. 2a). Zanda et al. (1993) found a wide diversity in the siderophile element contents of their metal grains, with only the average Co/Ni ratio corresponding to the solar value. The most convoluted metal grains analyzed in this study likely correspond to the coalescence and thus compositional averaging of many of these grains during chondrule-forming events (e.g. Zanda et al. 1993). In about 15 % of cases (see section 3.2), this averaging did not prevent anomalous compositions to arise, implying the contribution of relatively coarse precursor grains. The reduced compositional variability in isolated grains may be traced to their large sizes compared to other petrographic settings (see Fig. 3c), hence a greater averaging.

The volatile element depletion inferred for the chondrule precursors (for both the metal (this study) and the silicate (e.g. Jacquet et al. 2012) portions) would be naturally explained if they are the products of an *incomplete condensation* process, possibly in a protoplanetary disk context (Wasson and Chou 1974). This could in particular account for the paucity of sulfur (with 50 % condensation temperature of 664 K according to Lodders 2003) in type I chondrules (contrary to their counterparts in ordinary chondrites), as emphasized by Zanda et al. (2002), although a significant amount of S is now found in the fine-grained matrix (2.4 wt% on average according to Hutchison 2004). We note that the positive Ag anomaly suggests that Ag was actually less volatile than Cu, contrary to what literature calculations indicate (Wai and Wasson 1977; Campbell et al. 2003; Lodders 2003), but as the difference in 50 % condensation temperatures is rather slight (988 vs. 1029 K according to Campbell et al. 2003), this does not appear inconceivable. Our incomplete condensation scenario could explain why the most convoluted metal grains, i.e. those least affected by igneous reequilibration, tend to have solar Ni contents and Co/Ni ratios (see "Chemistry" subsection). We note that some amoeboid olivine aggregates in CR chondrites contain FeNi metal which Weisberg et al. (2004) found to be compositionally similar to margin grains in chondrules. The presence of anomalous patterns for refractory PGEs in some grains (Fig. 5a-c) suggests that CR chondrite metal preserves a record of fractionation processes at temperatures > 1400 K as well: precursors of PGE-enriched grains were presumably separated from the gas before condensation of main component siderophile elements, and the complementary PGE-depleted precursor grains were produced from further condensation of this residual gas. This fractionation may have been restricted to the source reservoir of a few precursor grains, or alternatively been general among the precursors of CR chondrule metal but essentially erased by averaging between PGE-enriched and PGE-depleted precursors. We note that refractory PGE (which are tightly intercorrelated, e.g. Fig. 6a) and main component siderophile element (e.g. Ni and Pd) tend to show some decoupling (around the solar ratio) for margin and isolated grains (Fig. 6b,c), specifically the refractory elements seem more variable than the main component — in contrast to interior grains which we will argue below are more processed. This would suggest that the refractory PGE were mostly carried by PGE-rich condensates and diluted in varying amounts in a moderately refractory metal, whose dominance, owing to solar abundances, would prevent large variations in Ni, Fe, Pd, Co in the resulting mixtures. That is, the high-temperature fractionation process described above may have indeed pertained to the bulk of CR chondrite metal, rather than to a handful anomalous grains.

Whether these precursors first formed as solid or liquid condensates, and whether they were molten upon aggregation (as "sprays of droplets" according to Hewins and Zanda 2012), has yet to be determined by a detailed study of the slightly melted objects (Zanda et al. 1993, 2002).

### 4.3 Metal-silicate equilibration and interior grains

We mentioned in the "Petrography" subsection 3.1 that interior grains tend to be rounded, suggesting that they solidified from molten droplets. This is at variance with the hypothesis of Wasson and Rubin (2010) that interior grains are unmelted relics of an earlier generation of Ni-richer grains. They had noted a chemical difficulty, namely that those grains would be *more* susceptible to melting than margin grains owing to their higher Ni content, but had invoked a fluxing of margin grains by FeS which would have lowered their melting point. However, measured S concentrations are low (mostly <<0.1 wt%) and similar for both petrographical settings. Textural evidence then leads us to conclude that interior grains (at least those that are rounded and Ni-rich, in highly melted chondrules) were molten and, far from being relics, were substantially equilibrated with their enclosing silicates. This is because diffusivities of siderophile elements in liquid metal (> $10^{-7}$ m²/s for Ni at 1750 K; Wasson and Rubin 2010), solid metal (>$10^{-14}$ m²/s for Co, Ni, Cu, Ru, Pd, Ir above 1400 °C; Righter et al. 2005) or silicate melt (>$10^{-14}$ m²/s for Re, Co, Fe above 1400 °C; Zhang et al. 2010) are sufficiently high for a 100 µm-diameter metal grain to be equilibrated for the slow cooling rates (<100 K/h) inferred from olivine trace element composition by Jacquet et al. (2012), or from the diffusion of Cu and Ga in coarse metal grains by Humayun (2012). The lack of zoning of interior grains observed by Lee et al. (1992) would be consistent with such an assumption.

Oxidation of Fe would account for the high Ni contents of interior grains, higher than that exhibited by metal grains in the least melted objects (3.5-6.9 wt%; Lee et al. (1992); Wasson and

Rubin (2010)), and their subsolar Co/Ni ratio, as Ni is more siderophile (in terms of the metal/silicate partition coefficient) than Co (e.g. Righter 2003). Oxidative loss of Fe would also be consistent with the relatively small size of interior grains (Fig. 3c). That even *type I* chondrule formation was accompanied by oxidation would reflect that, even for these objects, the estimated oxygen fugacities (e.g. Zanda et al. 1994; Schrader et al. 2013) are much higher than for a nebular gas (out of which the precursors would have presumably condensed ; see e.g. Grossman et al. 2012).

Further evidence for metal-silicate equilibration is found in the positive correlation between olivine/metal partition coefficient for Cr (which is dependent on oxygen fugacity) and the Fe content of olivine (Fig. 7a), as was first noted by Zanda et al. (1994); in contrast, P, which is more siderophile behaves coherently with Ni in the Ni-rich interior grains (Fig. 7b; see also Connolly et al. (2001) and Campbell et al. (2005)). The even more siderophile PGE (see Jones and Drake 1986; Capobianco et al. 1993) essentially preserve their (generally chondritic) pre-melting ratios in metal during equilibration since a very small fraction of it partitions into silicates (for silicate/metal partition coefficients much smaller than the mass fraction of metal (typically of order 20 %; see Ebel et al. 2008)). We note that measured Ni concentration in Renazzo olivine (73 ppm on average, with first and third quartiles at 43 and 143 ppm, respectively) reported by Jacquet et al. (2012) are somewhat in excess of equilibrium partitioning predictions (14 ppm for 7 wt% Ni in metal and a silicate/metal partition coefficient of $2 \times 10^{-4}$ (Jones and Drake 1986)); it is possible that some metal nuggets somewhat compromised the siderophile element section of these analyses, which focused on lithophile elements, although such nuggets were also monitored using GLITTER.

However, enrichment of Ni and other siderophile elements of metal is not accompanied by an increase of FeO content in olivine (Fig. 3d) as would be expected if this variation were driven by a *closed-system* oxidation of iron, as was noted by Wasson and Rubin (2010). This suggests an open-system scenario where Fe is first oxidized and thus enters the silicate melt, and is then transmitted to the surface and partially lost by evaporation. Fe evaporation was also suggested by Zanda et al. (2002) (see also Hewins et al. 1997) and would be supported by the fact that those chondrules with the most rounded interior grains show the most forsteritic olivine (Fig. 3e), similar to observations by Zanda et al. (2002). In a similar vein, some early loss of volatile P would be suggested by the fact that the Ni-rich interior grain branch of the P vs Ni plot (Fig. 7b) starts off at the *low-P* end of the distribution of the other grains, although recondensation of P on the latter might also account for the difference.

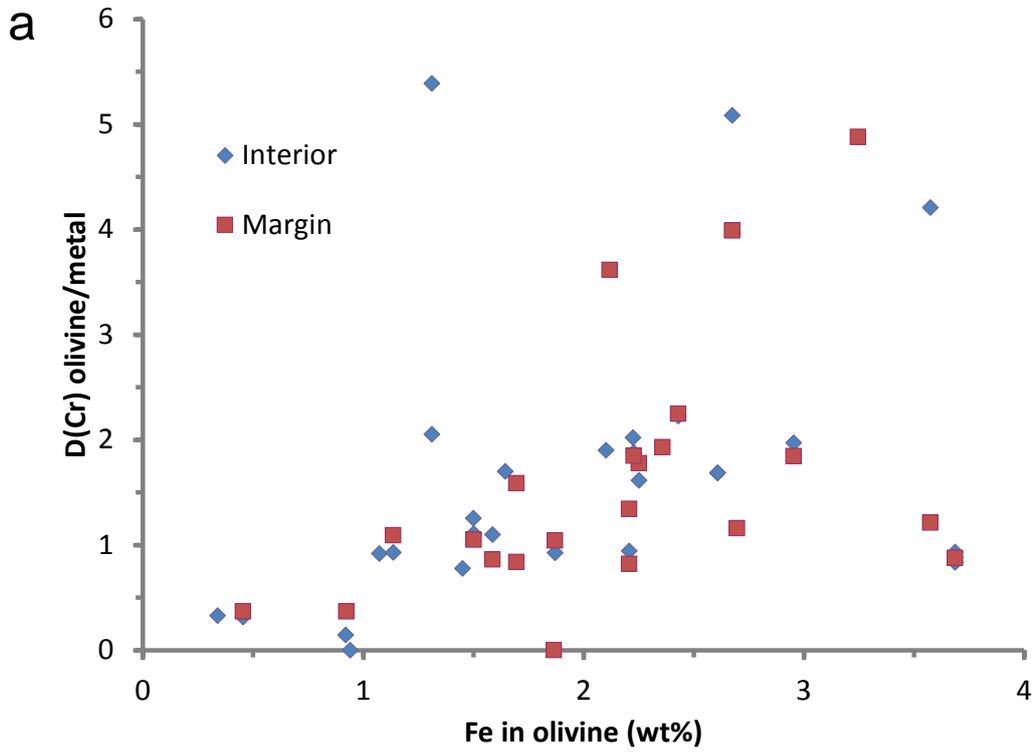

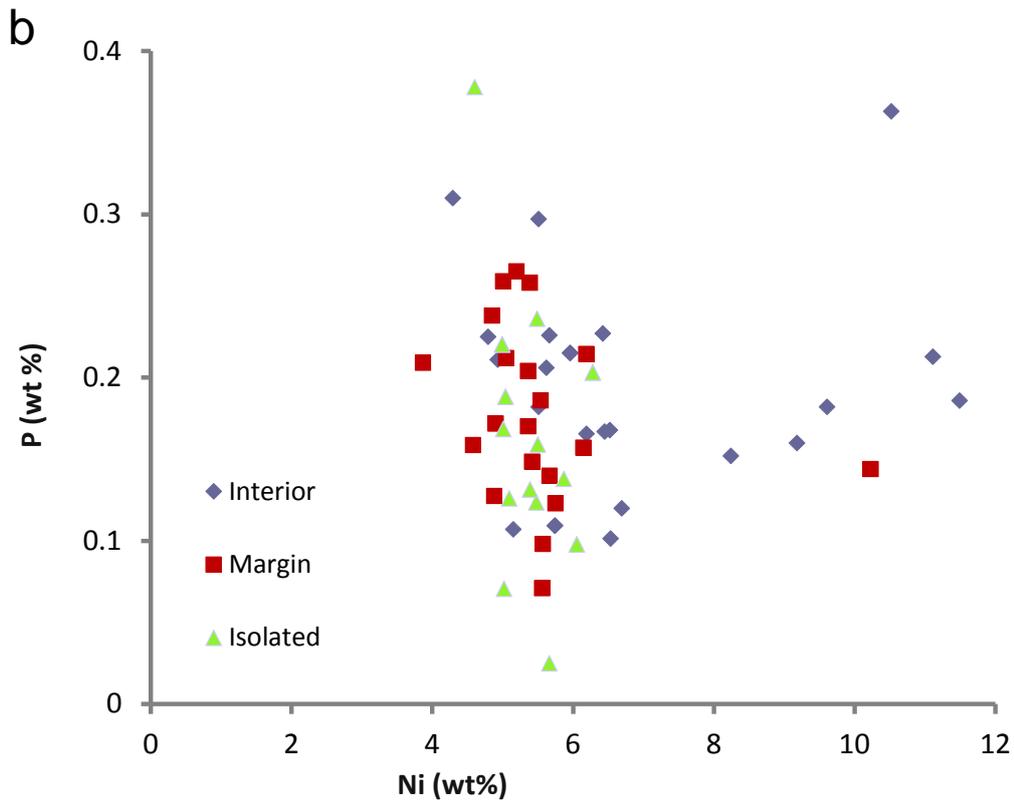

**Figure 7:** *(a)* Plot of olivine/metal partition coefficient for Cr as a function of Fe content of olivine. *(b)* Plot of P versus Ni in metal grains.

## 4.4 Formation of margin grains

The obvious depletion of metal in highly-melted chondrule interiors relative to their margin and to the least-melted objects is evidence (if significant recondensation onto chondrule surface is ruled out) for migration of interior grains toward the periphery, or even their escape from chondrules, which plausibly could both be due to the surface tension effects investigated theoretically by Uesugi et al. (2008), as proposed by Wasson and Rubin (2010). It seems however unlikely that highly-melted chondrules were ever covered by a continuous liquid film which beaded upon cooling, as envisioned by Wasson and Rubin (2010) or Humayun et al. (2010). First, we have never observed a continuous metal shell around a whole chondrule (unlike e.g. the continuous (silicate-laden) sulfide veneers around many ordinary chondrite chondrules (Hewins and Zanda 2012) or the kamacite-rich rims around some Kainsaz (CO3) chondrules described by Kring 1991) and even in 2D sections, margin grains seldom appear to have formed such a continuous shell around the host chondrule prior to an hypothetical beading. Second, the trace siderophile element heterogeneity in margin grains from single chondrules observed by Connolly et al. (2001) and confirmed by Humayun et al. (2010) is inconsistent with the rapid homogenization timescale that would be expected from diffusivities of siderophile elements in liquid metal (Wasson and Rubin 2010). Margin metal was thus likely not sufficiently abundant to constitute an interconnected shell, but formed discrete amoeboid grains by coalescence of smaller grains.

The morphological and compositional differences between interior and margin grains suggests that margin metal is not simply metal originated from the interior, although, as mentioned above, such outward migration certainly occurred and contributed to it. In fact, margin grains are more reminiscent of convoluted metal grains seen in medium melted chondrules (regardless of location) and appear to have been less heated than interior grains. Two further lines of evidence point toward the idea that the chondrule margin only record a brief thermal event: First, we have noted that metal margins were almost systematically included in the pyroxene-rich periphery of chondrules, and the monoclinic structure of the enstatite points to cooling rates of order 1000 K/h around 1300 K (Jones 1990). Second, Lee et al. (1992) noted frequent zoning of margin grains, with Ni decreasing from core to periphery, which could be explained by outside-in fractional crystallization during rapid cooling, as envisioned by Wasson and Rubin (2010) for large metal grains.

How can chondrule margins have been less heated than chondrule interiors? This appears conceivable only if the material precursor to chondrule margin was added (and briefly melted) *after* that precursor to the chondrule interior was first melted. The widespread occurrence, around type I chondrules, of igneous rims (see e.g. Fig. 1; Fig 2d) analogous to the slightly melted objects studied by Zanda et al. (1993, 2002) is evidence that such accretion of material was common in the chondrule-forming regions of CR chondrites. In some cases, multiple metal rings in single chondrules have been described (e.g. chondrule "Ch7" in Fig. 1 of Connolly et al. 2001; Fig. 1b of Wasson and Rubin 2010).

## 4.5 A scenario for CR chondrite chondrules

Piecing our different arguments together, we propose the following sequence of events to explain the formation of typical type I chondrules in CR chondrites, illustrated in Fig. 8:

(i) Aggregation of grains, or liquid droplets (previously produced during an incomplete condensation process).
(ii) A prolonged (> 1 day) high-temperature phase during which small metal grains coalesced to form rounded grains equilibrated with silicate (dominantly olivine), with evaporative loss of Fe from the chondrule. [Slightly and medium melted chondrules represent arrested stages of this process].
(iii) Accretion of a relatively fine-grained rim.
(iv) A rapid high-temperature phase that essentially affected the chondrule's periphery only, with formation of pyroxene (perhaps by addition of silica from the exterior, e.g. through the SiO gas as envisioned by Libourel et al. (2006)) and margin grains by melting of rim metal and interior grains.
(v) Further accretion of relatively fine-grained material (possibly heated *in situ* hereafter, if not molten upon deposition).

Written as such, this sequence makes no assumption on whether the different stages (i)-(v) were immediately adjacent/overlapping or not. It is conceivable, for instance, that stages (ii) and (iv) correspond to separate, independent thermal events (similar to the inference by Libourel et al. 2006), with the production of chondrule precursors and the igneous rim possibly representing further separate thermal episodes. But it appears possible, and perhaps simpler, to envision that stages (i)-(v) were part of a single heating episode, i.e. melting and aggregation may have taken place concurrently. In that case, stage (iv) would represent a "quenching phase", where the cooling rate would have accelerated, and the rim would represent the latest stage of dust agglomeration, which would have escaped thorough melting because of the falling temperatures. We have drawn a schematic temperature curve in Fig. 8 under this single-event assumption. One could think of stage (iv)-(v) as taking place as the chondrule was leaving the high-temperature region where it was melted in stages (ii)-(iii). Such a thermal history would be at variance with the shock wave model prediction, where the cooling rate is continuously *slowing down* after the temperature peak (e.g. Morris and Desch 2010).

The hypothesis of concurrent aggregation and melting would imply that the chondrule-forming region had high dust densities, similarly to what Rubin (2010) inferred for carbonaceous chondrites. To obtain a more quantitative estimate of that dust density $\rho_d$, we note that the thickness $e$ of a dust rim accreted during a time $t_d$ is related to it by:

$$\rho_s e = \Delta v \rho_d t_d, \quad (1)$$

with $\Delta v$ the velocity of dust grains relative to the chondrule and $\rho_s$ the internal density of the rim. $\Delta v$ is certainly a most uncertain parameter in the current ignorance of the astrophysical context of chondrule formation, but if we assume it to be of the same order of magnitude than the chondrule-chondrule relative velocity (as would be the case in incompressible hydrodynamical turbulence ; Ormel and Cuzzi 2007), the fragmentation velocity estimated at 1 m/s (Güttler et al 2010) could be taken as an upper bound. Then, for $t_d$ = 10 h, $e$ = 100 µm and $\rho_s$ = 3 x $10^3$ kg/m$^3$, we obtain a *lower* limit of $\rho_d$ = $10^{-5}$ kg/m$^3$, 3 orders of magnitude larger than the solid density expected in a Minimum Mass Solar Nebula model at 1 AU (Hayashi 1981). This is however comparable to the *chondrule* mass density (i.e. the number density times the mean chondrule mass) inferred from the frequency of compound chondrules. The latter may indeed be approximated by:

$$x_{cpd} = \frac{3\rho_c}{2\rho_s a} \Delta v \, t_{cpd}, \quad (2)$$

with $\rho_c$ the chondrule mass density, $a$ the chondrule radius and $t_{cpd}$ the time span where compound chondrules could form (the relaxation time of deformed chondrules, e.g. Rubin and Wasson 2005). Equations (1) and (2) may be combined in:

$$\frac{e}{a x_{cpd}} = \frac{2\rho_d t_d}{3\rho_c t_{cpd}}, \quad (3)$$

where $\Delta v$ has been eliminated. Adopting the compound chondrule frequency of CR chondrite Acfer 139 of 9.6 % (Hylton et al. 2005; higher than the 4 % measured in ordinary chondrites by Gooding and Keil 1981), if we take $t_{cpd} \sim t_d$, the "self-consistency" equation (3) imposes $\rho_c \sim \rho_d$ consistent (the large uncertainties notwithstanding) with the subequal fractions of chondrules and fine-grained matrix in CR chondrites. We note that the oxygen fugacity calculated by Schrader et al. (2013) for type I chondrules in CR chondrites also suggests an enhancement of the $H_2O/H_2$ ratio by 1-2 orders of magnitude over solar values. The large inferred density is also in line with those (>$10^{-2}$ kg/m$^3$) inferred by Alexander et al. (2008) for *ordinary* chondrites on the basis of Na retention (see also Hewins et al. 2012). The properties of CR chondrite chondrules would thus lend support to the emerging picture of localized chondrule-forming regions, highly non-representative of the average protoplanetary disk, at least in terms of temperatures and concentrations of solids.

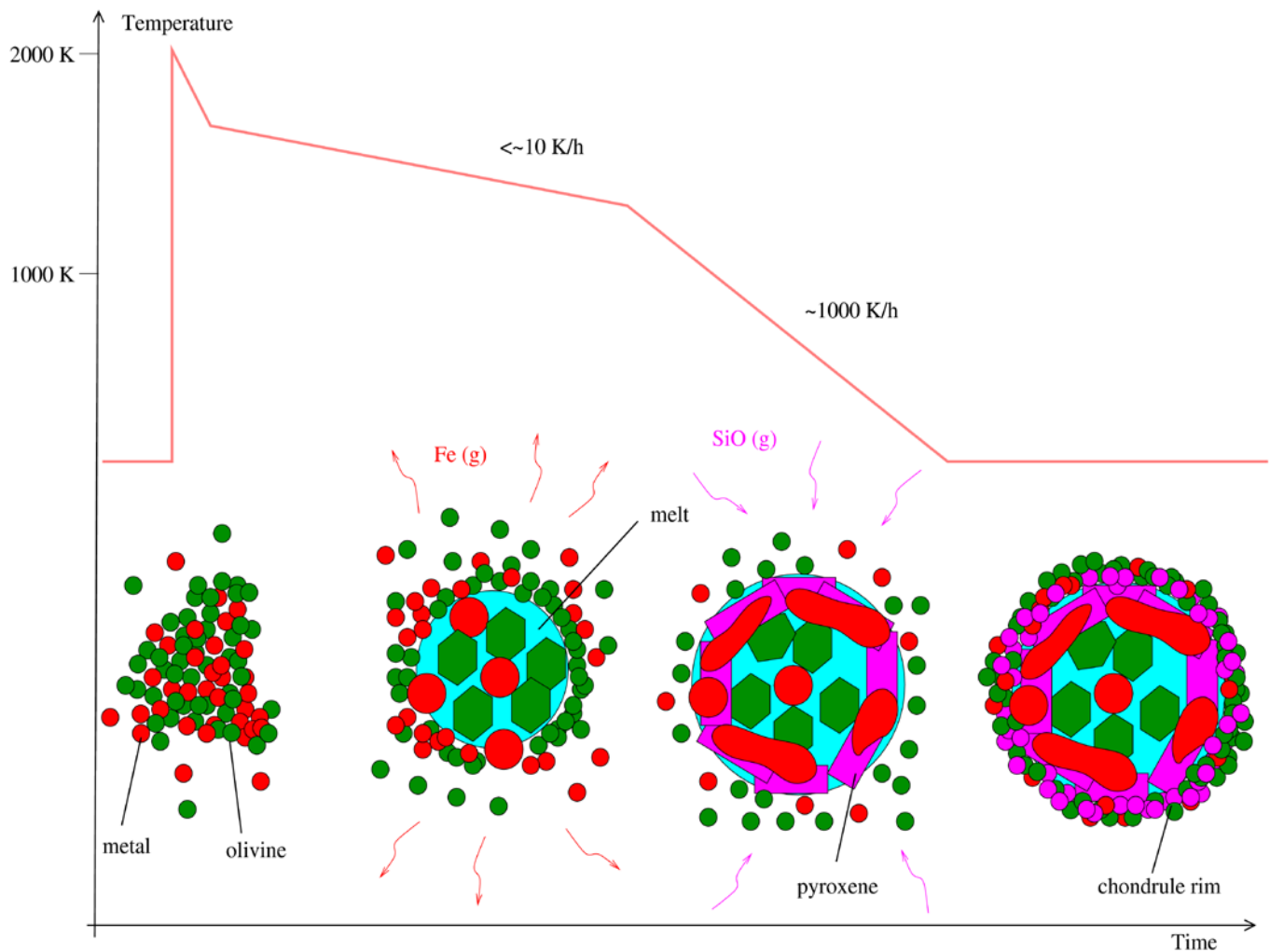

**Figure 8:** Sketch of the proposed formation scenario for highly melted type I chondrules in CR chondrites. An aggregate forms and is melted during a prolonged high-temperature event (with some Fe evaporation), hence the formation of large olivine crystals and metal nodules. Accretion of small grains on the molten chondrules continues. The temperature drops suddenly, leading to incomplete processing of the latest accreted metal grains; pyroxene formation is promoted by silica addition (e.g. from ambient SiO gas) apparently in that stage. Further accretion of small grains during cooling yields the outermost rim around the chondrule. A schematic temperature curve is also drawn. Olivine is color-coded in green, pyroxene in pink, metal in red and melt/mesostasis in cyan.

## *5. Conclusion*

We have analyzed trace element compositions of metal grains in ten different CR chondrites, distinguishing between interior, margin and isolated grains. We recover the positive Co-Ni correlation typical of CR chondrite metal, with the Ni-richest ones being rounded interior grains. In most (~85 %) grains, refractory siderophile elements are generally unfractionated relative to Ni, with volatile elements showing increasing depletion with decreasing 50 % condensation temperature. Averaged Ni-normalized abundance patterns for the different petrographic settings are essentially indistinguishable from each other.

The lack of systematic depletion of refractory siderophile elements in margin grains relative to interior grains argues against a condensation origin of the latter during chondrule formation. Likely, the different metal grains share a common, fine-grained precursor, probably best approximated by the least melted chondrules, which presumably owe their volatile element depletion to an incomplete condensation process, with likely a high-temperature fractionation between PGE-rich refractory condensates and gas that led to the condensation of PGE-poor metal, before mixing between these different precursor grains. Both interior and margin grains were molten during chondrule formation, but the former appear to have attained a higher degree of chemical equilibrium with the silicate than the latter (with in particular oxidation of Fe, subsequently lost by evaporation). This is understood as indicating that the chondrule margins record a later, rapid, high-temperature episode than that recorded by the chondrule interior, after more fine-grained material was accreted on the chondrule. It is possible that the two thermal episodes were parts of a single thermal event, with the rapid episode representing the "quenching phase" as the chondrule left its high-temperature region of formation, which would be at variance with a shock wave scenario. As dust accretion would have occurred concurrently with melting, this scenario requires high solid concentrations in the chondrule-forming regions, consistent with earlier constraints from compound chondrule frequencies or volatile element retention.

## Acknowledgments
We are grateful to the Programme National de Planétologie and the Institut Universitaire de France for their financial supports, NASA and Cecilia Satterwhite for the loan of the Antarctic meteorite sections, and the Muséum National d'Histoire Naturelle for that of the non-Antarctic samples. We thank Gretchen Benedix and Herbert Palme for their reviews which significantly improved the assessment and discussion of the data.